\documentclass[
 aps, pra,
 amsmath,amssymb,
 11pt,
 final,
tightenlines,
 twoside,
 twocolumn,
 nofloats,
nofootinbib,
 superscriptaddress,
showkeys,
showkeywords,
 ]
{revtex4-2}
\usepackage[T2A]{fontenc}
\usepackage[russian,english]{babel}
\usepackage{graphicx}
\usepackage{dcolumn}
\usepackage{bm}
\usepackage{longtable}
\usepackage{xcolor}
\usepackage[normalem]{ulem}
\usepackage{comment}

\usepackage{supertabular}
\input{maik.rty}
\usepackage{supertabular}

\setcitestyle{authoryear,round,aysep={,},citesep={;}}
\def\squareforqed{\hbox{\rlap{$\sqcap$}$\sqcup$}}

\def\sq{\ifmmode\squareforqed\else{\unskip\nobreak\hfil
\penalty50\hskip1em\null\nobreak\hfil\squareforqed
\parfillskip=0pt\finalhyphendemerits=0\endgraf}\fi}

\def\arcmin{\hbox{$^\prime$}}

\def\arcsec{\hbox{$^{\prime\prime}$}}

\def\utw{\smash{\rlap{\lower5pt\hbox{$\sim$}}}}

\def\udtw{\smash{\rlap{\lower6pt\hbox{$\approx$}}}}

\def\diameter{{\ifmmode\mathchoice
{\ooalign{\hfil\hbox{$\displaystyle/$}\hfil\crcr
{\hbox{$\displaystyle\mathchar"20D$}}}}
{\ooalign{\hfil\hbox{$\textstyle/$}\hfil\crcr
{\hbox{$\textstyle\mathchar"20D$}}}}
{\ooalign{\hfil\hbox{$\scriptstyle/$}\hfil\crcr
{\hbox{$\scriptstyle\mathchar"20D$}}}}
{\ooalign{\hfil\hbox{$\scriptscriptstyle/$}\hfil\crcr
{\hbox{$\scriptscriptstyle\mathchar"20D$}}}}
\else{\ooalign{\hfil/\hfil\crcr\mathhexbox20D}}%
\fi}}

\newcommand{\aap}{Astron. and Astrophys. }

\newcommand{\aaps}{Astron. and Astrophys. Suppl. }

\newcommand{\aj}{Astron.~J. }
\renewcommand{\apj}{Astrophys.~J. }
\newcommand{\apjs}{Astrophys.~J. Suppl. }

\newcommand{\azh}{Astron.~Zh. }

\newcommand{\mnras}{Monthly Notices Royal Astron. Soc. }
\newcommand{\pasa}{Publ. Astron. Soc. Australia }

\newcommand{\pasp}{Publ. Astron. Soc. Pacific }

\renewcommand{\nat}{Nature }

\newcommand{\memras}{Memoirs of the Royal Astronomical Society }

\begin{document}

\selectlanguage{english}

\title{RADIO SOURCES OF THE SURVEY ON THE DECLINATION OF THE PULSAR IN  CRAB NEBULA (Dec = $+22^\circ$)}

\author{\firstname{A.~A.}~\surname{Kudryashova} }
\affiliation{Special Astrophysical Observatory of the Russian Academy of Sciences, Nizhny Arkhyz, 369167 Russia}

\author{\firstname{N.~N.}~\surname{Bursov} }
\affiliation{Special Astrophysical Observatory of the Russian Academy of Sciences, Nizhny Arkhyz, 369167 Russia}

\author{\firstname{S.~A.}~\surname{Trushkin}}
\affiliation{Special Astrophysical Observatory of the Russian Academy of Sciences, Nizhny Arkhyz, 369167 Russia}

\begin{abstract}
The results of the analysis of 205 brightest sources ( $S>15$ mJy), which were found in the sky survey at the declination of the pulsar in the Crab Nebula, are presented. The survey was conducted at a frequency of 4.7~GHz using a three-beam radiometer complex installed in the focus of the Western Sector of the RATAN-600 radio telescope in 2018-2019.
Based on the measurements and data collected in the database of astrophysical catalogs CATS built radio spectra of objects. For a quarter of all detected sources, data at a frequency higher than 4~GHz were obtained for the first time, and for the rest, they were appended. The variability of radiation sources on the scales of the year, from days to
months, was studied. The greatest change in the radio flux was found in the blazar B2~1324+22. The search for daily variability was carried out for 26 the brightest sources with an average value of $S_{4.7} \sim 250$ mJy. All sources are identified with objects from optical and infrared catalogs. Radio luminosity is calculated for 112 objects with a known redshift.
\end{abstract}

\maketitle

\section{1. INTRODUCTION}

Radio surveys of the sky are the main research method that allows to obtain new information about a large number of objects of various types. Among the most famous radio reviews are 3C~\citep{1959MmRAS..68...37E}, 4C~\citep{1965MmRAS..69..183P, 1967MmRAS..71...49G}, Parkes~\citep{1991PASA....9..170O}, GB6~\citep{1996ApJS..103..427G},  TGSS~\citep{2017A&A...598A..78I}, GLEAM~\citep{2017MNRAS.464.1146H}, VLASS~\citep{2020PASP..132c5001L}.
A significant contribution to the study of radio sources was made by surveys of large areas of the sky at the frequency \mbox{1400 MHz} on the interferometer VLA, NVSS~\citep{1998AJ....115.1693C} and FIRST~\citep{1994ASPC...61..165B}.
Single telescopes with a large aperture also play an important role: 500-m FAST, 100-m GBT, as well as RATAN-600, which provide new information in a wide frequency range.

Several large surveys of the sky were carried out on the RATAN-600 radio telescope: the Zelenchuk survey in the declination range $0^\circ~-~14^\circ$ ~\citep{1992AZh....69..225A}, in-depth review “Cold” at a frequency of 3.9 GHz, as a result of which an RC catalog was obtained ~\citep{1991AandAS...87....1P, 1993A&AS...98R.391P, 1996BSAO...42....5B, 2010AstBu..65...42S}, a multi-frequency round-the-clock survey of the two-degree near-zenith region (RZF, \mbox{$\delta = 41\,.\!\!^\circ5\pm{1^\circ}$})) for ten years at the declination of the bright source 3C 84 \citep{2004GrCo...10....1P}. Since 2017, daily sky surveys at different declinations have been conducted in the Western Sector of RATAN-600 using a complex of four-channel radiometers at a central frequency of 4.7 GHz. In the observations, the antenna temperature of the sky is measured depending on the average sidereal time when a section of the sky passes through the  beam pattern (BP) of the telescope.

Here we discuss the results of a study of a sample of sources found in the survey at the declination of one of the brightest objects in the sky in the radio band of the Crab Nebula. The detected sources were identified by 
cross-identification with a large amount of radio data using the CATS data base (\citep{CATS}) in order to build  radio spectra of the sources. As a result of processing, the values of the  flux densities at a frequency of 4.7 GHz were obtained. Using these data we refined and refilled the radio spectra of the sources, analyzed the light curves of the sources at different time scales during the year and estimated the variability of the flux density. For the sample sources a study was conducted on base of the available data. To do this, we used the astrophysical databases NED
\footnote{\tt https://ned.ipac.caltech.edu/}, \mbox{Roma-BZCAT}\footnote{\tt https://www.ssdc.asi.it/bzcat5/}, VizieR\footnote{\tt https://vizier.cds.unistra.fr/} and their tools. Most of the sources are identified with the SDSS optical catalog (DR16)\footnote{\tt http://cas.sdss.org/dr16/en/home.aspx}, as well as with GAIA (DR3)\footnote{\tt https://gea.esac.esa.int/archive/}, Guide Star Catalog\footnote{\tt https://outerspace.stsci.edu/display/GC}   and Pan-STARRS\footnote{\tt https://catalogs.mast.stsci.edu/panstarrs/}.

 \section{OBSERVATIONS AND DATA-PROCESSING}

The observations were carried out with the Western sector (azimuth$270^\circ$  ) antenna of RATAN-600 with the antenna elevation of the pulsar in the Crab Nebula (\mbox{($\delta = +22^\circ00\arcmin52\arcsec$)}). The sky view strip had an area of more than 200 square degrees. The estimated height at which the antenna of the Western Sector was focused was 
$H_0 = 32^\circ48\arcmin47\arcsec$. 

The survey was conducted throughout the year from May 28 2018 to May 30, 2019 with a complex of three similar radiometers at a central frequency of 4.7 GHz ($\lambda = 6.38$ cm) and with a bandwidth of 600 MHz divided into channels with a width of 150 MHz each \citep{2018AstBu..73..494T}. The receiving horns of the three radiometers were positioned along the focal line of the secondary mirror in such a way that the second horn was in the focus of the radio telescope. The data  acquisition system was described by \citet{2011AstBu..66..109T}. The survey data was divided into hourly of local sidereal time records (drift scans) of observations. The methods of data-processing these records are described by  \citet{BursovPhD}.

In the acquisition system, the sampling (due to the frequency of the quartz oscillator) was initially 490 microseconds for primary aim to search for fast radio bursts (FRB), then it was 245 microseconds and 61 microseconds for  the solution of giant pulses at the pulsar in the Crab nebula. For data-processing the usual cosmic radio sources  we  increased  the  signal to noise ratio (S/N) the records were compressed to sampling
of  0.097~s. At the same time this value itself changed slightly due to temperature variations. To achieve uniformity of the data, their interpolation was carried out to 0.1~s. In the channels of each radiometer, the records were aligned to the average value. They were measured and averaged, which led to a decrease in noise on recordings by about half.

For each calendar month of observations, about 25 high-quality daily records of each hour of observations were obtained with each radiometer. Thus 12 average  records were obtained during the year. The sensitivity at the output of the radiometers at a frequency of 4.7 GHz reached  \mbox{3~мК\,с$^{-1}$}, and with
an effective area of the Western Sector of about 1000 m$^2$ the sensitivity of the telescope
is about 5 mJy/beam. This value is limited by atmospheric variations and industrial interference, as well as the effect of confusion for the an BP having angular dimensions of $1\arcmin\times35\arcmin$.

Details of passing sources through the BP of the radio telescope antenna in an azimuth of $270^\circ$ with a relatively large size of the secondary mirror of the fifth type and an asymmetric opening of the antenna of the Western Sector are described in  \citep{Mayorova}, where it is shown that this does not lead to a change in the vertical size of the BP shape at 4.7 GHz. To identify the detected sources  with the sources from the NVSS catalog, the coordinates of the latter were recalculated by the ephemerides routine {\tt svp} \citep{2012AstBu..67..225K}. The identification was carried out by the cross-identification method taking into account the properties of the sources and their  angular shift relative to the center of the beam pattern  $dH$, where $dH =  H_0-H_{\rm src}$ dH. An important part of the processing was the binding of records in sidereal time and amplitude, which were unstable.

\begin{figure}[h]
\includegraphics[width=7.8cm]{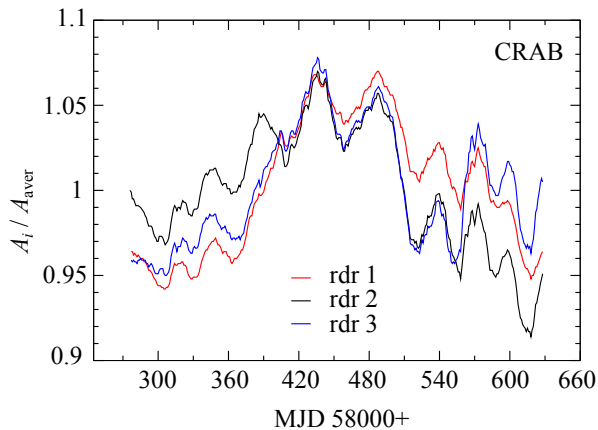}
\caption{ Amplitude variations ($A$) of the Crab Nebula in the survey on the Western sector of RATAN-600 at a frequency of 4.7~GHz in relative unit.}
\label{ris:clb}
\end{figure}

The time instability was primarily due to network failures and electricity surges. In the absence of calibration signals, changes in ambient temperature and the radiometers themselves, which operated in a warm mode without thermal stabilization, affected the signal amplitude level. Another factor of instability was the change in the state of the radiometers themselves over time (electricity, microwave amplifiers, etc.). Therefore, in operation, the amplitude of the sources was measured not in antenna temperatures ($T_a$), but in units of measured amplitude (A). However, the amplitudes of the source records were measured in the antenna temperature units (mK). The antenna temperatures of the bright Crab Nebula, which corresponds to the primary standards, were used for the general calibration of the observation records by sensitivity during the year. Changes in the antenna temperature of the source are usually associated with a change in the gain of the radiometers during the entire measurement cycle. The calibration curves obtained from the Crab Nebula are shown in 
\ref{ris:clb}.

\begin{figure*}[t]
\includegraphics[height=5.0cm]{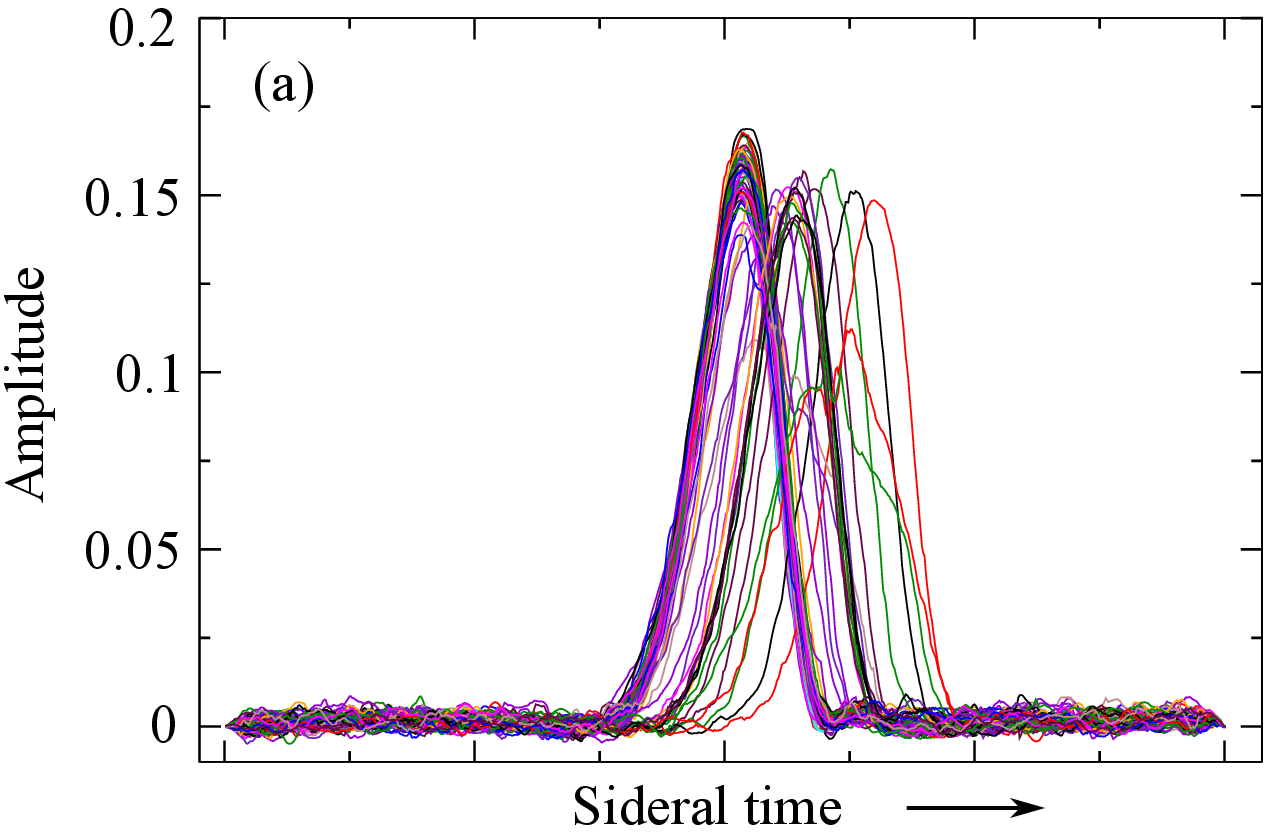} 
\includegraphics[height=5.03cm]{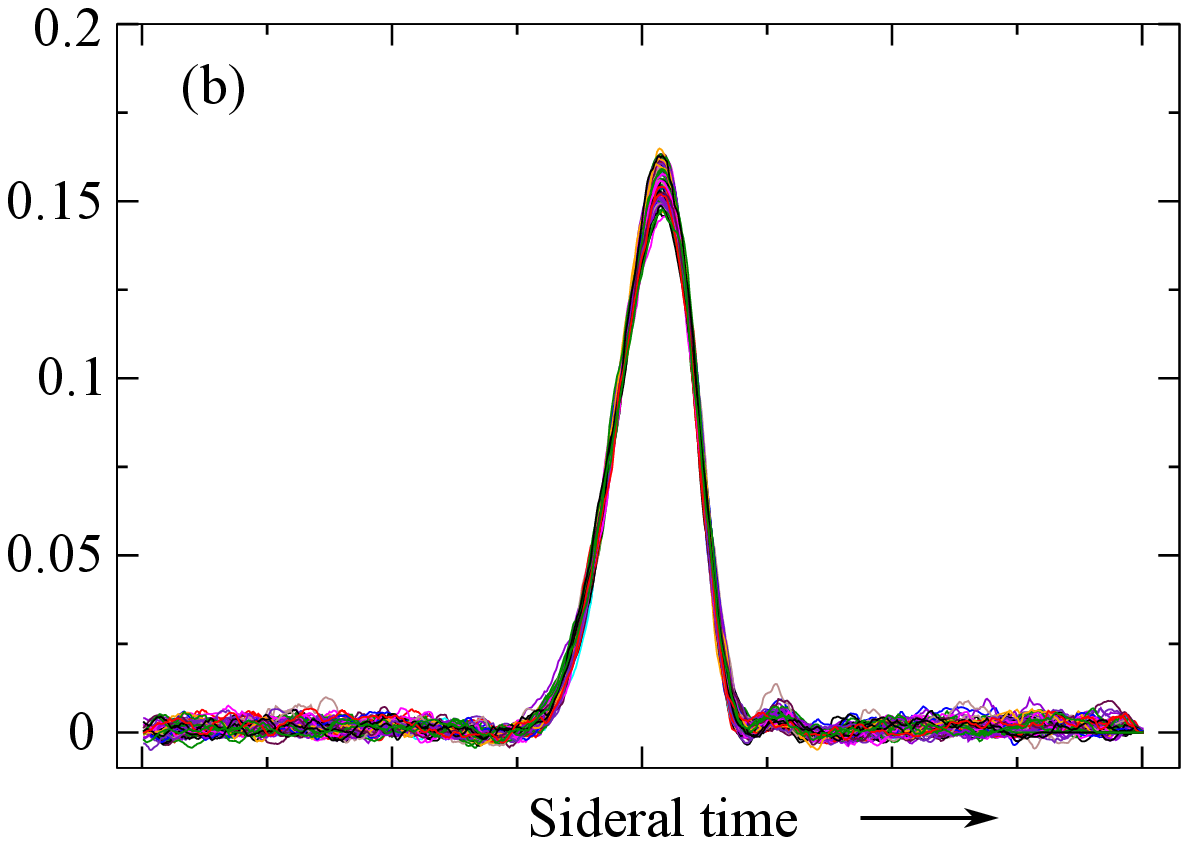}
\caption{Source NVSS\,J003147$+$215347: (a)~--- initial, interference-free recordings for the year of observations; (b)~--- after correction by ephemerides and amplitude
}
\label{ris:corr}
\end{figure*}

When accumulating records, it would be necessary to take into account the change in the position of the sources during the year due to precession and nutation relative to a fixed antenna. Corrections were also made to the mean sidereal time due to the failures of GPS-transmitter binding. In the first case, the solution to the problem was to shift the records according to the calculated ephemerides. Since not all the charts had bright sources, the general time binding of records was carried out using a weighted average curve with a general error at a level not exceeding $1-2$~s. This had a slight effect on the accuracy of the position of the sources and their amplitude when averaging records, since the dimensions of the half-width of the beam pattern in terms of half-power (HPBW) were greater than 5~s. Nevertheless, in order to increase the accuracy of measurements of the source parameters (sidereal time, amplitude and half-width), sources from hourly observation records were cut out with a window of about 1 minute of time and their position was corrected in accordance with their ephemerides.

Example of correction for a source NVSS J003147+215347 (hereinafter, for the sake of brevity
of the names in the tables and figures, we  we will omit the NVSS name) from the middle $A = 150$ mK is shown in Fig. 2, where panel (a) shows interference—free records for a year of observations; panel (b) - after correction with accurate ephemerides, with an amplitude measurement accuracy of about 10\%; panel (c) — after correction for ephemerides and amplitude, with an accuracy of 3.5\%. 
For sources with $A$ is about 10 mK, the average error in determining $A$ after correction is about 10\%. When forming a sample of sources, we averaged the adjusted records for a year of observations. To achieve the maximum signal-to-noise ratio to these records, convolution with the real DN obtained from the source was applied. The criterion for including sources in the sample of bright objects was \mbox{$S/N > 50$} with the amplitude of the sources  \mbox{$A \ge 5$}mK. As a result, for the entire sample of sources located no further than  $|dH| < 10\arcmin$ from the focal line of the antenna bottom, the minimum flux density was 15 mJy To link sources by current density depending on their shifts from the focus 
a subsample of bright sources with power-law spectra was compiled, according to which calibration curves were constructed: the ratio of the flux density of each source at the viewing frequency ($S_{4.7}$) to its amplitude ($A$) in
depending on the shifts of the source from the center of the each beam. Figure 3 shows a histogram of the distribution of the spectral density of the source flux at 4.7~GHz. Only one source in the sample has a flux density of   $S_{4.7} > 1$Jy. Despite the fact that the sources are bright — with a high signal—to-noise ratio - the flux density for half of the sources  $S_{4.7} < 50$ mJy. The inset shows calibration curves for the flux density, depending on the flux, 
 $S_{4.7}/Ta \sim f(dH)$, where $Ta\approx A$.

\begin{figure*}[t!]
\centering
\includegraphics[width=12cm]{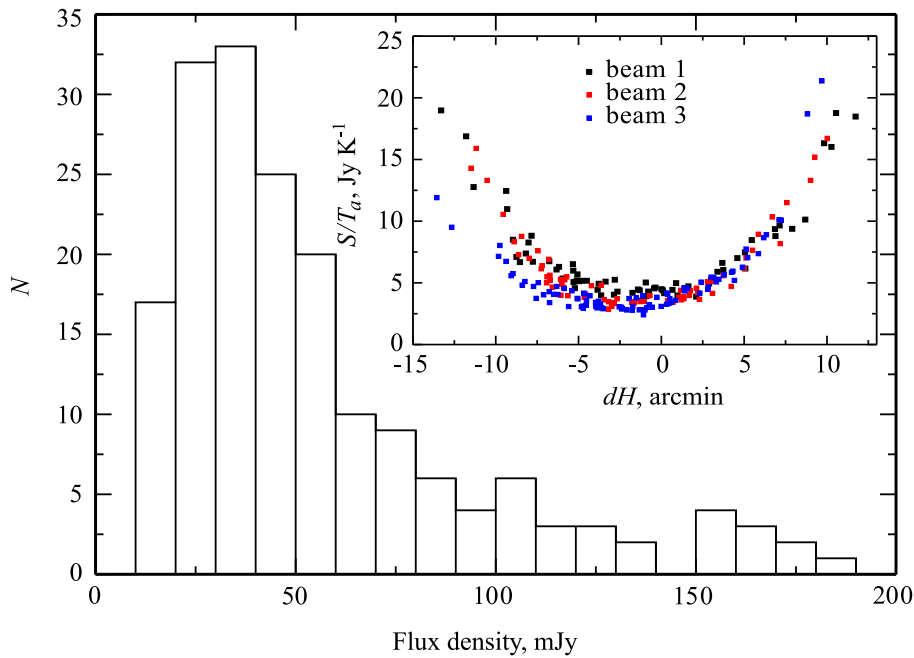}
\caption{A histogram of the distribution of the sampling flux density of 205 sources at a frequency of \mbox{4.7~GHz}. The inset shows the calibration curves for the flux density for three radiometers.}
\end{figure*}

\section{RADIO SPECTRA}

In the review by \citet{2017ApJ...836..174C}, sources with different spectral energy distribution were studied: sources with a peak at a frequency of about 1 GHz (Gigahertz Peaked Spectrum, GPS), compact sources with an ultra-steep spectrum (Compact Steep Spectrum, CSS) and sources with a peak at high frequencies (High Frequency Peaked, HFP), which represent a class of radio - emitting active galactic nuclei (AGN), which are believed to be the young precursors of massive AGN, such as Centaurus A and Cygnus A \citep{1998PASP..110..493O, 1991ApJ...380...66O, 2005AandA...432...31T, 2010MNRAS.408.2261K}
The spectra of GPS and HFP sources have a noticeable peak and decrease in fluxes on both sides of the peak. CSS sources are believed to have properties similar to GPS sources and HFP, but their peak frequencies are significantly
the lower \citep{2000AandA...361L..25P}. Therefore, the main difference between GPS, CSS and HFP sources is the frequency of the spectral peak and the size of the object. GPS and HFP sources They have linear dimensions of about 1 kpc and
peak frequencies of about 1-5 GHz and more than 5 GHz, respectively \citep{2000AandA...363..887D}. While CSS sources have a linear size of about 1-20 kpc and are generally considered to have the lowest peak frequencies, up to hundreds of megahertz. With the CATS database the sample sources were identified with sources of other radio catalogs at various frequencies. According to the review and literature data, radio spectra of the studied objects were constructed. The 
{\tt spg} program from the {\tt FADPS} package was used for this\citep{1993BSAO...36..132V}. As the calculation showed, the data on Frequencies above 4~GHz were obtained for the first time for a quarter of the sources, and the spectra were supplemented for the rest of the source sample. The spectral index was calculated as the tangent of the angle of inclination of the straight line on the graph  $\log\nu \sim \log S_{\nu}$ between the NVSS data on 1.4 GHz and the data obtained in this work at 4.7 GHz. The error of the spectral index was calculated using well-known formulas. The 16 sources in our review have lower frequencies of maximum flux densities 1 GHz, which follows, as a rule, from the measurements of the low-frequency GLEAM survey. 

\begin{figure*}[t]
\includegraphics[width=0.445\textwidth]{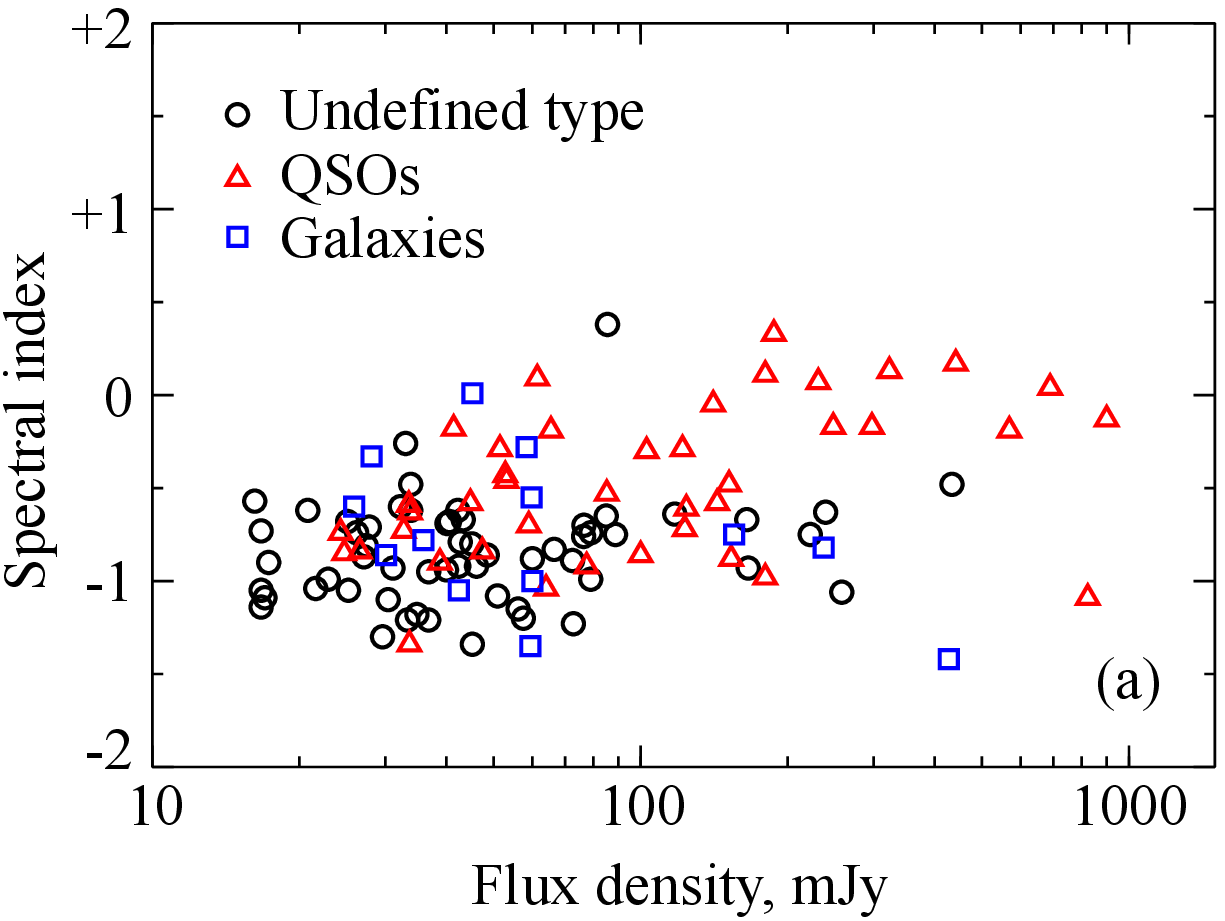}
\vspace*{0.2cm}
\includegraphics[width=0.45\textwidth]{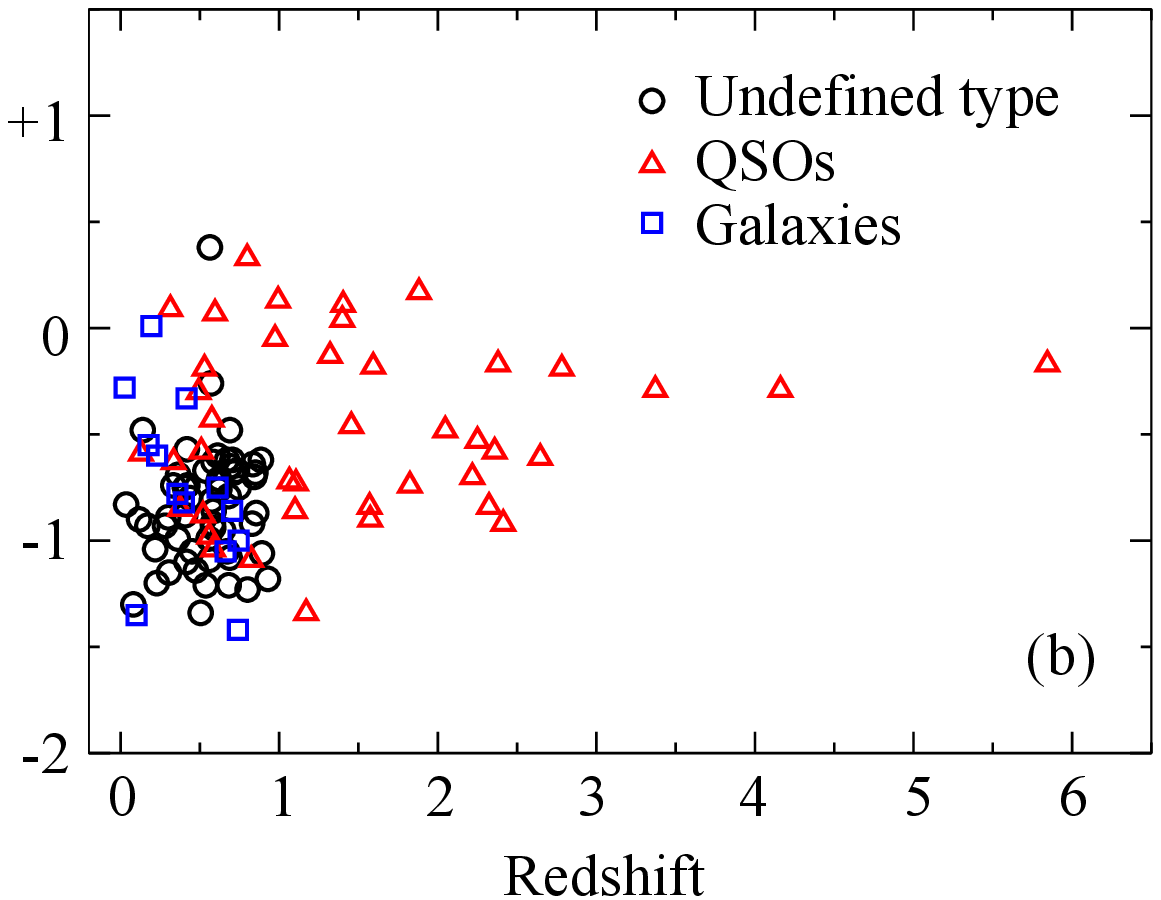}
\label{ris:z-spi}
\caption{ The dependence of the spectral indices of 205~ sources at the frequency \mbox{4.7~GHz} on the flux density (a) and the redshift $z$ (b). The empty circles show sources of unknown type, triangles and squares~--- quasars and galaxies, respectively.
}
\label{ris4:spi}
\end{figure*}

Theses objects are candidates for CSS sources. Five sources with a maximum within frequencies 1-2 GHz  are candidates for GPS sources. The peak frequency of NVSS J072614+215319 is 8.5 GHz, and it most likely belongs to HFP sources.
USS (Ultra steep spectrum) sources are defined as compact radio sources with ($ \alpha < -1.1$ (\mbox{$S_{\nu}\sim \nu^{\alpha}$}). They were given attention when searching for large-z AGN, as it was found that the steep spectra correspond to the highest redshifts \citet{2001MNRAS.327..907J, 2006MNRAS.366...58D, 2006MNRAS.371..852K, 2009MNRAS.394.2197B}. The increasing  of the spectral indices of sources at low frequencies with an increase in the flux density was given by  \citet{1982AandA...113..150G, 1993ApJ...405..498W, 2012MNRAS.421.1644R}, as well as by 
 \cite{2023A&A...675L...3D} for 43,000 VCSS catalog sources at 340 MHz. 
\cite{2018AstBu..73..142Z}  constructed a dependence at 3.94 GHz for 304 and 396 radio sources of the RC catalog, in which the inverse dependence was obtained – flattening of spectral indices, which is mainly characteristic of the contribution of distant and bright quasars. The same result was obtained in this work at a frequency of 4.7 GHz. 
(Fig.~\ref{ris4:spi}a)). According to our work, the “redshift – spectral index” dependence also shows a flattening of the spectral indices of quasars with increasing $z$ (Fig. \ref{ris4:spi}b). However, in the work of Laing and Peacock (1980) at 178 MHz, with  $S_{178} \ge 10$ Jy, spectra became steeper with increasing redshift for galaxies with powerful radio emission and steep spectra. 

\begin{figure}[t]
\includegraphics[width=0.48\textwidth]{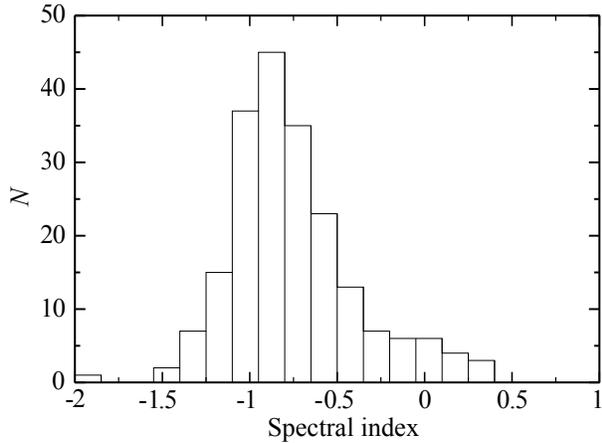}
\caption{ Histogram of the distribution of spectral indices of sampling sources at the frequency \mbox{4.7~GHz}.
}
\label{ris:spi}
\end{figure}

This is another class of objects from the USS, High-z radio galaxies (HzRGs) — galaxies with active star formation at large z. Some of the discovered galaxies of this work, apparently, also belong to this class of objects. The distribution of spectral indices is shown in Fig.\ref{ris:spi}, its peak value $\alpha \approx -0.9$. The classification showed that 124 sources (61\%) have a normal the power-law spectrum (\mbox{($-1.1<\alpha<-0.5$)}), 22 (11\%) are sources with a peak in the spectrum, 25 objects  (12\%) are sources with a very steep spectrum, 26 sources (13\%) have a flat spectrum 
(\mbox{($-0.5 < \alpha < 0.3$}), three sources have an inverted spectrum ($\alpha > 0.3$), and eight sources have a low—frequency fall-down. 

The spectra of 205 sources are shown in Appendix A, where the blue filled circles indicate the data from CATS data base, and the red ones indicate the measured values of the 4.7~GHz flux density according to the data from this work. The spectra show from one to three measurements, depending on the quality of the data and the shifts of sources from the focus of three radiometers.

\section{ VARIABILITY OF RADIO SOURCES}

For the first time, the variability of extra-galactic source was found in the bright quasar 3C 273, in the age-old optical measurements of which minor variations in brightness were detected by \citep{1963Natur.198..650S}. The first evidence of the variability of radio sources was published in the works on the source CTA 102 at a wavelength of 32.5 cm 
(Sholomitskii, 1965) and on the source 3C 273 (\citep{1965Sci...148.1458D}). A model of variability in synchrotron self-absorption in an adiabatically expanding source was described by \citet{1966Natur.211.1131V}. To date, the study of the variability of radio sources by various methods has become the subject of a wide range of works. 

Long-term monitoring of the variability of objects of various types is carried out with RATAN-600. The search for long–term variability and variations of radiation on a scale of one day or more for sources at declinations 
4$^\circ$--6$^\circ$ and 10$^\circ$--12$\,.\!\!^\circ5$ for a long time (20 and 10 years of observations, respectively) was carried out by the radio laboratory of the Moscow State University Traffic Police (\citep{2008ARep...52..278G, 2013ARep...57..344G}). A number of works are devoted to the study of the flash activity of microquasars. 
\citet{2023AstBu..78..225T} discovered the evolution of the spectrum of Cygnus X-3 flare radiation on a time scale less the orbital period (4.8h) of the binary system. \citet{2022AstBu..77..246S} considered the variability of the sample of galaxies with sources of hydroxide (OH) megamaser radiation measured with RATAN-600.

In this work, we are guided by two approaches to the study of the variability of sample sources, considering the daily variability during the year for a subsample of the brightest sources (quasars and blares) and estimating the index of variability when accumulating data for each month of observations during the year.

\subsection{ Search for the variability of weak sources}

We performed calculation of the index of variability on the scale of the year for the total sample of sources. The data were averaged in all cases for each month of observations. As a result, a light curve was constructed for each sample source from 12 average amplitudes and their errors for the entire year of observations. All
the data obtained were adjusted for a change in the gain along the light curves of the Crab Nebula calibration source, followed by Gauss analysis (Fig. ~\ref{ris:clb}). We calculated the dimensionless index of variability $I_{\rm var}$ using the modified formula from \citet{1992ApJ...399...16A}, where instead of the values of densities The amplitudes of the fluxes are used. This replacement allows us to estimate the index of source variability depending on the signal-to-noise ratio.
\begin{equation}
\label{eq:Var}
 I_{\rm var}=\cfrac{(A_{\rm max}-\sigma_{A_{\rm max}})-(A_{\rm min}+\sigma_{A_{\rm min}})}
{(A_{\rm max}-\sigma_{A_{\rm max}})+(A_{\rm min}+\sigma_{A_{\rm min}})},
\end{equation}

where $A_{max}$ and $A_{min}$ are the maximum and minimum values of the average amplitudes for all months of observation; 
$\sigma_{A_{\rm max}}$, and  $\sigma_{A_{\rm min}}$ are amplitude measurement errors.

In a third fraction of the sources,  $I_{\rm var}$ was determined from observations in two or three beams. For analysis, data with an axial value of $A$ was selected for sources passing closer to the center of the bottom of the input horn of the radiometer. The calculated values of the source variability index are summarized in the Table \ref{BursovA1_part}.
For fainter sources, the detected variability may be related to measurement inaccuracy at the level of relatively high noise ($S/N \sim 10$--$30$). Therefore, for sources with  $A < 10$(30\%), the index was evaluated with an accumulation of records for up to two months. Finally, for a group of sources with an amplitude of less than 6 mK (15\%), accumulation was carried out with an interval of three months. Figure 6 shows the calculated $I_{\rm var}$ values, where there is only one relatively large value of $I_{\rm var}$ = 0.3. This is the well-known blazar NVSS J132700+221050 (B2 1324+22) at $z$ = 1.4 with a change in flux density more than twice during the year. Five more sources have a radiation variability with an
$I_{\rm var} = 0.2$ with an amplitude change of one and a half times. These are the sources NVSS J114417+220752, NVSS J141726+220539, NVSS J160203+220931, NVSS J170744+220049 and NVSS J174005+221100. For the rest of the sources, the 
$I_{\rm var} < 0.15$, which means that 97\% of the sources do not have noticeable variability on a one-year scale.
\renewcommand{\baselinestretch}{0.72}
\setlength{\tabcolsep}{3.5pt}
\begin{table*}
\caption{\label{BursovA1_part}
Data of observations. The columns are  
 (1)~--- Name; (2)~--- number of the the closest beam; (3)~--- mean amplitude of signal $A$; (4)~--- variability index 
$ I_{\rm var}$; (5)~--- flux density in NVSS catalog  $S_{1.4}$; (6)~--- flux density in this survey $S_{4.7}$, (7)~--- spectral index with error $\alpha_{4.7}$; (8)~--- shift of a source from focal line $dH$; (9)~--- width of source $HPBW$; (10)~--- number of measurements (months) $N$. It is fraction of the  total table, which are accessible as Additional materials} \medskip
\begin{tabular}{c|c|r|r|r|r@{$\,\pm$\,}l|c|r|r|c}
\hline
\multirow{2}{*}{NVSS name} & \multirow{2}{*}{Beam} &   \multicolumn{1}{c|}{\multirow{2}{*}{$A$}} &   \multicolumn{1}{c|}{\multirow{2}{*}{$ I_{\rm var}$}}  &  \multicolumn{1}{c|}{$S_{1.4}$,} &  \multicolumn{2}{c|}{$S_{4.7}$,} &  \multirow{2}{*}{\multirow{1}{*}{$\alpha_{4.7}$}} &   \multicolumn{1}{c|}{$dH$,} &   $HPBW$,  &  \multirow{2}{*}{$N$} \\
	  &    &        &       &  mJy  &           \multicolumn{2}{c|}{ mJy }      &               &  arcmin & \multicolumn{1}{c|}{s}     &    \\
\hline
(1)&(2)&(3)&(4)&\multicolumn{1}{c|}{(5)}&\multicolumn{2}{c|}{(6)}&(7)&\multicolumn{1}{c|}{(8)}&\multicolumn{1}{c|}{(9)}&(10) \\
\hline
  J000727$+$220413    &  1  &   9.9 &  0.14 &  816.2 &   180.0 & 11.8  &     $-$0.75\,$\pm$\,0.12 &  11.3 &   12.6 &  12 \\
 J001145$+$215912    &  1  &  21.1 &  0.05 &  302.2 &   123.1  &  6.6  &     $-$0.74\,$\pm$\,0.06 &   3.3 &    8.7 &  12 \\
 J002130$+$215319    &  3  &   8.1 &  0.02 &   53.6 &    24.3  &  3.1  &     $-$0.64\,$\pm$\,0.09 &   0.2 &    8.7 &   6 \\
 J002337$+$215624    &  1  &  39.3 &  0.08 &  442.6 &   166.1  & 6.8  &     $-$0.81\,$\pm$\,0.06 &  $-$1.2 &    8.1 &  12 \\
 J003147$+$215347    &  3  &  78.0 &  0.06 &  777.4 &   258.2  & 9.3  &     $-$0.86\,$\pm$\,0.09 &   0.9 &    8.1 &  12 \\
 J004157$+$215423    &  2  &  10.3 &  0.02 &   55.5 &    32.2  &  2.5 &     $-$0.45\,$\pm$\,0.13 &  $-$1.3 &    9.5 &  12 \\
 J011852$+$215144    &  2  &   7.8 &$-$0.01 &  124.2 &    40.6 &  2.7 &     $-$0.92\,$\pm$\,0.07 & $-$5.8 &   10.9 &  12 \\
 J012428$+$215454    &  2  &   5.8 &  0.11 &   46.1 &    20.8  &  1.7 &     $-$0.63\,$\pm$\,0.06 & $-$0.8 &    8.9 &  11 \\
 J012729$+$215136    &  3  &   5.4 &$-$0.05 &   56.9 &    17.0 &  2.7 &     $-$0.99\,$\pm$\,0.08 & $-$2.9 &    8.6 &   6 \\
 J013352$+$220125    &  1  &   5.5 &  0.11 &  113.1 &    53.1  &  4.1 &     $-$0.62\,$\pm$\,0.08 &   6.5 &   10.8 &  12 \\
\hline
\end{tabular}
\end{table*}

\begin{figure} []
\includegraphics[width=0.47\textwidth]{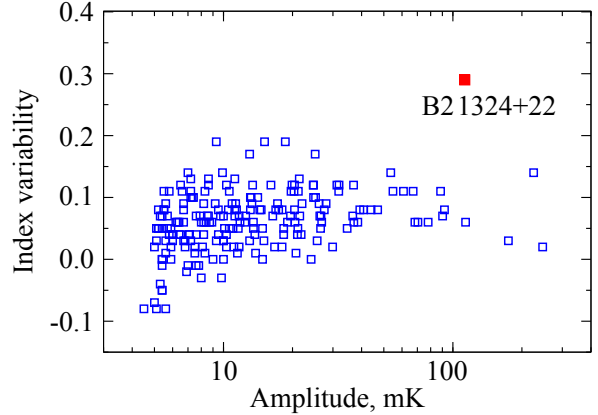}
\caption{ The indices of variability (in dimensionless units) for all 205 sources on \mbox{4.7~GHz}. Blazar B2\,1324$+$22 by $z = 1.4$ with a brightness change of more than two times is shown separately}
\label{ris:ivar}
\end{figure}

\subsection {Search for the variability of bright sources}

\begin{figure}[h]
\includegraphics[width=0.49\textwidth]{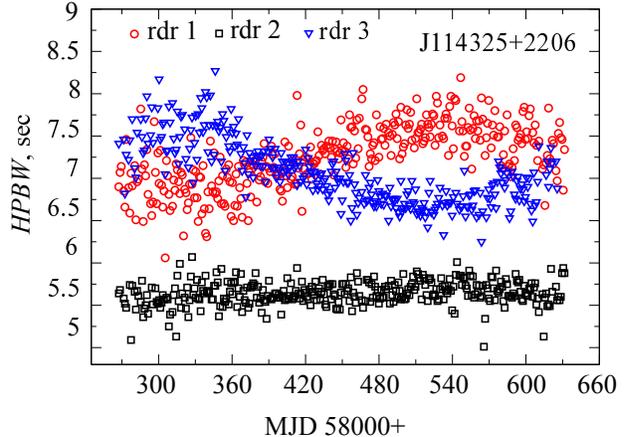}
\caption{ Changing the half-width of the source NVSS\,J114325$+$220656 when passing through the BOTTOM of the input horns of three radiometers (rdr\,1,...,3), one of which was in focus (the second).
}
\label{ris:1143}
\end{figure}

For the 26 brightest sources, which are relatively close to the central section of BP, more accurate amplitude measurements were carried out to search for a change on the scales of several days. The optimal accuracy of the result was the summation of data for three days of observations at each point of the light curve.
Initially, the study of variability During the year, observations were carried out according to their amplitude. However, due to precession, the source changes its position relative to the central section of the bottom rays. So some of the changes in $A$ were related to this fact, and not to the variability (Fig. \ref{ris:1143}).

Therefore, the search for the variability of sources was carried out by their integral radiation. The integrals of the sources on each record of observations were determined as the product of the sum of the amplitudes at each point
P is discrete in time, according to the formula  \mbox{$Int=\sum_{i=1}^{n}(\Delta{t}*A_i$)}. To eliminate the non-Gaussian noise from interference, a polynomial fitting was inserted into the light curves of each source using the least squares method, relative to which the noise threshold values of $3\sigma$ were estimated with the removal of exceeding values. 
Then the light curves of the sources were adjusted for seasonal temperature changes in the gain of the radiometers.
Figure 8 shows the light curves quasars from a sample of the brightest 26 sources. As mentioned above, the indices of variability of these objects, except for B2 1324+22, have small values, from which it can be concluded that, taking into account errors in determining the flux density, their radiation level has changed no more than one and a half times.

\begin{figure*}
\centering
\includegraphics[scale=1]{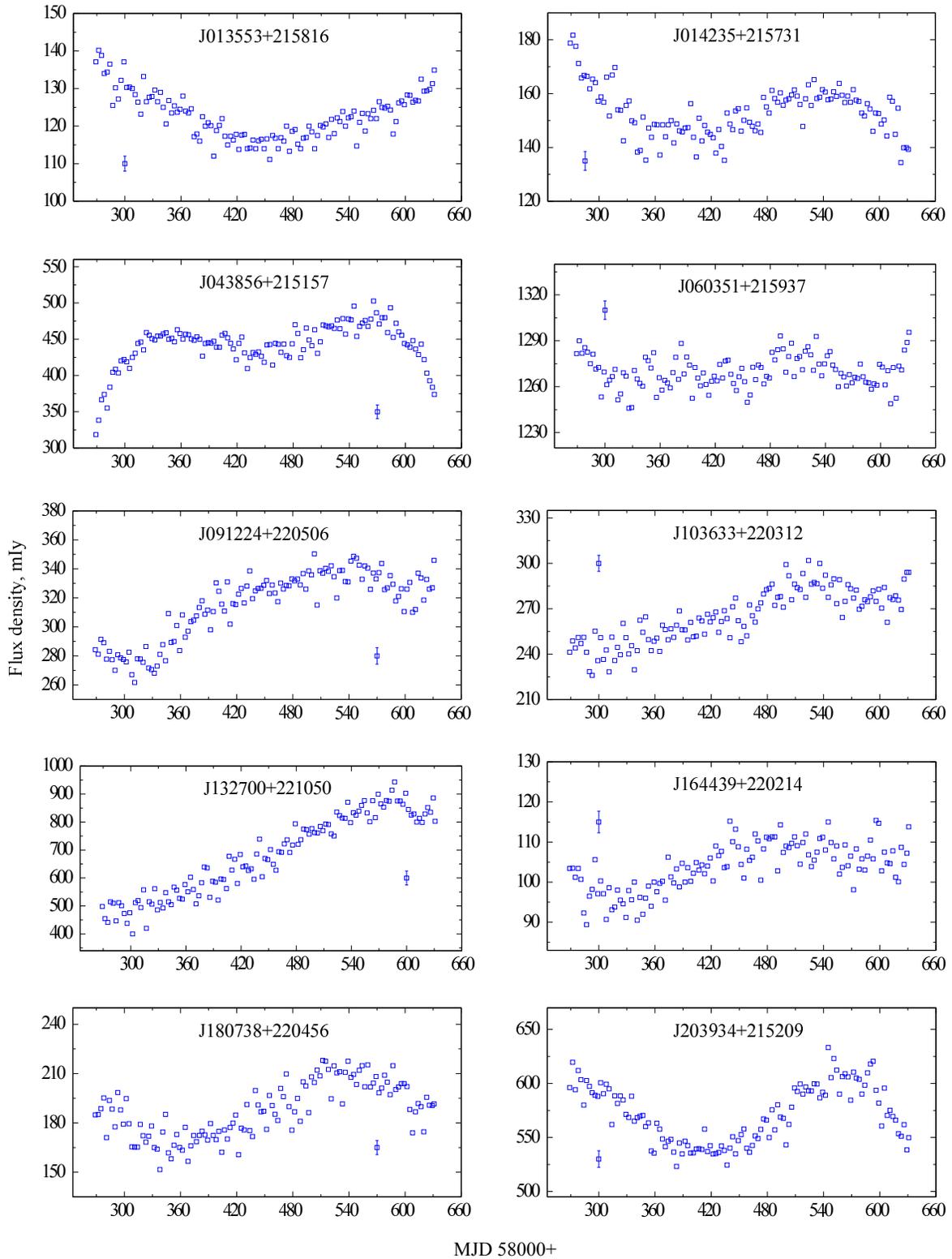} 
\caption{The daily variability of the flux density of quasars and blazars at the frequency \mbox{4.7~GHz}. The NVSS catalog name is omitted for brevity. }
 \label{ris:curve}
\end{figure*}

In Fig.~\ref{ris:curve} the light curves of flat-spectrum quasars and blazars from a sample of the brightest 26~sources are presented. As mentioned above, the indices of variability of these objects, except B2\,1324$+$22, have small values, from which it can be concluded that, taking into account errors in determining the flux density, their radiation level has changed no more than one and a half times. Nevertheless, the figures show noticeable changes in the brightness of the radiation throughout the year. So, three sources, namely: NVSS\,J014235$+$215731, NVSS\,J180738$+$220456~and~NVSS\,J203934$+$215209, probably show a periodicity on a scale of about one year. 

Three more sources: NVSS\,J091224$+$220506, NVSS\,J103633$+$220312 and NVSS\,J164439$+$220214, show an increase in flux density. At the source NVSS\,J013553$+$215816, a decrease in the radiation level was observed, followed by an increase after six months of observations. The source NVSS\,J043856$+$215157 had two brightenings during the year. The first occurred 80~ days after the start of the review, the second~---200~days after the first. At the brightest source of the view ($S_{1.4} > 1$~Jan) NVSS\,J060351$+$215937 (4C\,$+$22.12), candidate for blazars \citep{2014ApJS..215...14D} with a power spectrum \mbox{($\alpha = -0.7$)}, no changes in the radiation level were detected.

\section{OPTICAL AND INFRARED IDENTIFICATIONS}

The literature data available for the sample sources were analyzed. To this end, using the method of cross-identification within a radius of $15^{\prime\prime}$, we identified with the sources of the VizieR, NED, Roma-BZCAT astrophysical database. As a result, information was obtained on the physical characteristics of objects: their optical magnitudes, redshifts and classes of objects (quasar, blazar, galaxy, etc.).

In the optical range, most sources are \mbox{(147; 72\%)} identified with the SDSS~(DR16) directory.
Since the SDSS survey does not cover the entire sky, an additional identification was carried out with the optical catalogs Gaia\,DR3, Guide Star Catalog, Pan-STARRS. 174~ sources (85\%) have been identified with infrared sources in the WISE catalog.
Data on stellar magnitudes in the optical range, morphological characteristics (extended/point), and photometric redshifts for extended~--- are taken from the SDSS catalog. Spectroscopic data are available for some objects in the catalog, in which case the physical characteristics of the objects (their class: galaxy or quasar) and their spectroscopic redshifts are known. Identification with the GAIA catalog of extragalactic objects, DR3
Identification with GSC outside the SDSS coverage area gave morphology for 48~objects.
Additionally, in the Pan-STARRS catalog, outside the SDSS coverage area, two sources are identified with empty fields.
A total of 11 sample objects are in empty optical fields.
Three sources are located in optical fields with a large number of optical sources in the Pan-STARRS catalog. Due to the high density of objects in the images and due to the lack of information about their nature, it was not possible to unambiguously identify the sources of the sample under study that fall into this field. The results of the identification are given in the Table~\ref{BursovA2}.

In the presence of several measurements of the redshifts of objects, the spectroscopic method was preferred for further analysis. The three sources have similar measurements of spectral redshifts in different independent studies. For the NVSS\,J091914$+$22051 source, two values of the spectroscopic redshift are given in the Table~\ref{BursovA2}, since they differ significantly.

As a result, all sources were identified with optical and infrared catalogs.
Only 30\% of the sources have a certain physical type.
From the entire sample: quasars~--- 18\%, blazars~--- 6\%, galaxies~--- about 6\%. Point objects of indeterminate type make up ~ 28\% of the entire sample, extended ~--- 34\%, empty fields~--- 6\%. Five objects have been identified only with infrared data. For 57\% of the identified objects, redshift, spectroscopic or photometric, is known.

\LTcapwidth=0.99\textwidth
\renewcommand{\baselinestretch}{0.7} \setlength{\tabcolsep}{1.5pt}
\begin{longtable*}{c|c|c|cl}
\caption{ Cross-identification of sampling sources with optical catalogs.
The columns indicate: (1)~--- the name of the source; (2)~--- redshift, marked s~--- spectroscopic, p~--- photometric; (3)~--- the physical type of the object, according to generally accepted abbreviations:
QSO~--- quasar, BL\,Lac~--- blazar, FSRQ~--- quasar with a flat radio spectrum, G~--- galaxy, BClG~--- large galaxy cluster, GinCl~--- galaxy in a cluster, GinGroup~--- galaxy in a group; (4)~--- optical morphological type of object: Pt~--- point, Ext~--- extended, VisS~--- source in a field with a high density of optical objects.
References: [1]~--- The Roma-BZCAT, [2]~--- SIMBAD, [3]~--- NED, [4]~--- Gaia\,DR3 Extra-galactic, [5]~--- SDSS DR16, [6]~--- Pan-STARRS, [7]~--- Guide Star Catalog}\medskip \label{BursovA2}\\
\hline
\multirow{2}{*}{ NVSS name}  & \multirow{2}{*}{Redshift}  & \multirow{2}{*}{Class}  &   \multicolumn{2}{c}{Optical} \\
  &   &  &    \multicolumn{2}{c}{type} \\  \hline
(1) & (2) & (3) & \multicolumn{2}{c}{(4)}  \\
\hline
\endfirsthead
\caption{(Continued} \medskip\\
\hline
\multirow{2}{*}{ NVSS name}  & \multirow{2}{*}{Redshift}  & \multirow{2}{*}{Class}  &   \multicolumn{2}{c}{Optical} \\
  &   &  &  \multicolumn{2}{c}{type} \\  \hline
(1) & (2) & (3) &  \multicolumn{2}{c}{(4)}\\ \hline
\endhead
\hline
\endfoot
\endlastfoot
  J000727$+$220413 &  0.559 s [4]    &  QSO\,[3]       &    Ext& [5]    \\
  J001145$+$215912 &  1.064 s [4]    &  QSO\,[4]       &    Pt &[5]    \\
  J002130$+$215319 &  1.824 s [5]    &  QSO\,[2,5]     &    Pt &[5]     \\
  J002337$+$215624 &  0.169 s [5]    &  --             &    Ext& [5]   \\
  J003147$+$215347 &  0.891 p [5]    &  --             &    Ext& [5]    \\
  J004157$+$215423 &  0.612 p [5]    &  --             &    Ext& [5]   \\
  J011852$+$215144 &  --             &  --             &    EF & [5]    \\
  J012428$+$215454 &  0.671 s [5]    &  --             &    Ext& [5]   \\
  J012729$+$215136 &  0.560 p [5]    &  --             &    Ext& [5]    \\
  J013352$+$220125 &  1.455 s [5]    &   QSO\,[2,5]    &    Ext& [5]  \\
  J013553$+$215816 &  3.372   [3]    &   QSO\,[2]      &    Pt & [5]    \\
  J013756$+$215459 &  --             &  --             &    Ext& [5]   \\
  J014235$+$215731 &  2.047 s [5,\,4]&  QSO\,[5]       &    Pt&  [5]    \\
  J015218$+$220707 &  1.321 s [5]    &   BL\,Lac\,[2]  &    Pt&  [5]   \\
  J015641$+$215651 &  0.646 p [5]    &   --            &    Ext& [5]    \\
  J020007$+$215700 &  --             &  --             &    Pt&  [5]   \\
  J020434$+$215328 &  0.843 p [5]    &  --             &    Ext& [5]    \\
  J021018$+$215908 &  --             &  --             &    EF& [5]    \\
  J022110$+$215551 &  0.575 p [5]    &  --             &    Ext& [5]    \\
  J022754$+$215451 &  0.100 p [5]    &   Galaxy[2]     &    Ext& [5]   \\
  J023001$+$215304 &  0.581 p [5]    &  --             &    Ext& [5]    \\
  J023004$+$215909 &  0.529 s [4]    &   BL\,Lac[2]    &    Pt&[7]    \\
  J023349$+$215317 &  0.300 p [5]    &   --            &    Ext& [5]    \\
  J024928$+$215441 &  --             &   --            &    EF&[6]    \\
  J030252$+$215513 &  --             &   --            &    Pt&[7]     \\
  J031619$+$215555 &  --             &   --            &    Pt&[7]   \\
  J032313$+$215724 &  --             &   --            &    VisS&     \\
  J033524$+$215521 &  --             &   --            &   EF&[6]   \\
  J035142$+$215749 &  --             &   --            & Pt&[7]      \\
  J035933$+$215457 &  --             &   --            & Pt&[7]     \\
  J040036$+$215408 &  --             &   --            & Pt&[7]      \\
  J040755$+$215100 &  --             &   --            & Pt&[7]        \\
  J041555$+$215800 &  --             &   --            & Pt&[7]      \\
  J041913$+$220304 &  0.363 p [5]    &   --            & Ext&       \\
  J042211$+$220241 &  0.885 p [5]    &   --            & Ext&        \\
  J043312$+$215529 &  --             &   --            & EF&        \\
  J043458$+$215540 &  --             &   --            & Pt&[7]      \\
  J043507$+$215511 &  --             &   --            & Pt&[7]     \\
  J043856$+$215157 &  0.140 p [5]    &   --            & Ext& [5]       \\
  J044655$+$215448 &  --             &   --            & Pt& [5]       \\
  J045004$+$215814 &  0.280 p [5]        & --         & Ext & [5]  \\
  J052948$+$215521 &  0.280 p [5]    &   --            & Ext& [5]       \\
  J053431$+$220101 &  --             &   Crab          & Crab&      \\
  J060019$+$220715 &  --             &   --            & Pt& [5]        \\
  J060351$+$215937 &  --             &   BL\,Lac\,[2]      & Pt& [5]       \\
  J060640$+$215939 &  --             &  --             & Pt&[7]      \\
  J062240$+$215752 &  --             &   --            & Pt&[7]     \\
  J062250$+$220025 &  --             &  --             & Pt&[7]      \\
  J062727$+$220051 &  --             &  --             & Pt&[7]     \\
  J063017$+$215630  &  --            &   --            & Pt&[7]       \\
  J063101$+$215642  &  --            &   --            & Ext& [7]     \\
  J063446$+$220640  &  --            &   --            & Pt& [5]        \\
  J063727$+$220237  &  --            &   --            & Pt& [5]       \\
  J064711$+$215825  &  --            &  --             & Pt& [5]        \\
  J065630$+$220308  &  --            &   Galaxy\,[2]     & Pt& [7]     \\
  J070714$+$220459  &  --            &   --            & VisS&       \\
  J072300$+$215925  &  --            &   --            & Pt& [7]     \\
  J072319$+$220100  & 0.176 p [3]    &   Galaxy\,[3]     & Pt& [7]      \\
  J072351$+$220241  & 0.128 s [4]    &    QSO\,[4]       & Pt& [7]     \\
  J072543$+$220352  & --             &   QSO\,[4]       & Pt& [7]      \\
  J072614$+$215319  & 1.882 s [4,\,3]  &  FSRQ\,[1]      & Pt& [7]     \\
  J072820$+$215306  & 5.844 s [4]    & QSO\,[4]          & Pt& [7]      \\
  J073556$+$220848  & --             &  --             & Pt& [5]       \\
  J074642$+$220024  & --             &  --             & EF& [5]        \\
  J075706$+$215424  & --             &  --             & Pt& [5]       \\
  J081212$+$220024  & 1.106 s [5]    &   QSO\,[2,3,4,5]  & Pt& [5]       \\
  J081725$+$215840  & --             &  --             & Ext& [5]      \\
  J082353$+$220041  & --             &  --             & Pt& [5]        \\
  J085037$+$220615  & 1.570 s [5]    &   QSO \,[2]       & Pt& [5]       \\
  J090614$+$220010  & 0.520 s [5]    &   Galaxy \,[2]       & Ext& [5]       \\
  J091224$+$220506  & 0.993 s [5]    &   QSO \,[2,3,4]   & Pt&  [5]      \\
  J091914$+$220519  & 1.55 s [5]     & QSO\,[5]          & Pt&  [5]      \\
		    & 2.25 s [4]     &                 & & \\
  J092045$+$220433  & 0.034 s [2]    &   GinGroup\,[2]  & Ext& [5]      \\
  J092601$+$220136  & 0.622 p [5]    &  --             & Ext& [5]       \\
  J094736$+$220136  & 0.570 p [5]    &  --             & Ext& [5]      \\
  J094836$+$220053  & 0.716 s [5]    &  QSO[5,2,3]      & Ext& [5]       \\
  J101104$+$220805  & --             &  --             & Pt& [5]       \\
  J101401$+$215825  & 0.305 p [5]    &  --             & Ext& [5]       \\
  J102016$+$220940  & 0.314 s [5]    & QSO\,[5]         & Ext& [5]      \\
		    & --             & BL\,Lac\,[2]      & &\\
  J102154$+$215930  & 0.740 p [5]    & RadioG\,[2]      & Ext& [5]      \\
  J102408$+$220347  & 0.140 p [5]    &  --            & Ext& [5]      \\
  J103633$+$220312  & 0.595 s [5]    & BL\,Lac\,[2]       & Pt& [5]        \\
  J103943$+$215743  & 0.612 s [5]    &    Galaxy\,[5]   & Ext& [5]      \\
  J104254$+$220127  & 0.705 p [5]    & --              & Ext& [5]      \\
  J104702$+$221033  & 0.035 p [5]    &   STAR\,[2]      & Ext& [5]      \\
  J105234$+$220602  & --             &  --             & VisS& [5]     \\
  J105430$+$221055  & 4.161 s [4]    & BL\,Lac\,[1,2]     & Pt& [5]       \\
		    & --             & Galaxy\,[4]      &  &     \\
		    & --             &   QSO\,[4]       &  &     \\
  J105435$+$220011  & --             & --              & EF&        \\
  J110025$+$220156  & 0.621 p [5]    & --              & Ext& [5]      \\
  J110323$+$220337  & --             & --              & EF &        \\
  J112119$+$215947  & --             & --              & Ext& [5]       \\
  J112829$+$220729  & 0.691 p [5]    & --              & Ext& [5]       \\
  J113033$+$215728  & 0.399 s [5]    & Galaxy\,[5]      & Ext& [5]       \\
  J114325$+$220656  & 0.824 s [3]    & QSO\,[3]         & Ext& [5]       \\
  J114417$+$220752  & 0.575 s [5]    &   BClG\,[2]      & Ext& [5]       \\
  J114821$+$220825  & 0.800 p [5]    &  --             & Ext&  [5]      \\
  J115311$+$220654  & 0.415 p [5]    &  --             & Ext&  [5]      \\
  J121156$+$220455  & 0.117 p [5]    &  --             & Ext&  [5]      \\
  J125433$+$221103  & 0.509 s [5]    & BL\,Lac\,[2]       & Pt&  [5]       \\
		    & --             & Galaxy\,[5]      &    &         \\
  J130253$+$220758  & 0.332 p [5]    & --              & Ext& [5]       \\
  J130651$+$221119  & 0.421 p [5]    & --              & Ext& [5]       \\
  J131128$+$220306  & 0.475 p [5]    & --              & Ext& [5]       \\
  J132700$+$221050  & 1.398 s [5]    &  BL\,Lac\,[2]      & &     \\
		    & --             & FSRQ\,[1]        & Pt& [5]        \\
  J132749$+$220503  & 1.100 s [5]    &   QSO\,[2,4,5]    & Pt& [5]        \\
  J133212$+$220549  & 0.505 p [5]    &  --             & Ext& [5]       \\
  J133629$+$221033  & --             &  --             & Pt &[5]        \\
  J133928$+$220822  & 2.323 s  [5]   &   QSO\,[2,5]     & Pt& [5]        \\
  J135116$+$221110  & 1.574 s  [5]   &   QSO\,[2,4,5]   & Pt &[5]        \\
  J135313$+$220540  & 0.615 p  [5]   &  --             & Ext& [5]        \\
  J140808$+$220155  & 0.587 p  [5]   &  --             & Ext &[5]       \\
  J141242$+$215939  & 0.408 p  [5]   &  --             & Ext& [5]       \\
  J141351$+$220647  & 0.360 p  [5]   &  --             & Ext &[5]       \\
  J141619$+$220840  & --             &  --             & EF &        \\
  J141726$+$220539  & 0.563 p  [5]   &  --             & Ext &[5]       \\
  J143106$+$220505  & 0.713 p  [5]   &  --             & Ext &[5]       \\
  J143249$+$220759  & --             &  --             & Pt &[5]        \\
  J144057$+$220142  & 0.080 p  [5]   &  --             & Ext &[5]       \\
  J144924$+$221206  & --             &  --             & Pt&[5]      \\
  J150123$+$221122  & 0.688 p  [5]   &  --             & Ext &[5]       \\
  J151105$+$220806  & 0.586 p  [5]   &  --             & Ext &[5]       \\
  J151319$+$220255  & 0.850 p  [5]   &  --             & Ext& [5]       \\
  J151830$+$220313  & 0.681 p  [5]   &  --             & Ext &[5]       \\
  J152224$+$215808  & --             &  --             & EF &        \\
  J153343$+$220725  & --             &  --             & Ext& [5]       \\
  J153652$+$220207  & 0.230 p [5]    &   Galaxy\,[2]    & Ext& [5]       \\
  J154535$+$220400  & 0.218 p [5]    &  --             & Ext& [5]       \\
  J154631$+$215741  & 0.734 s [3]    &  --             & Ext& [5]       \\
  J155354$+$215927  & 0.745 s [3]    &  --             & Ext &[5]       \\
  J155513$+$215939  & --             &  --             & Pt &[5]        \\
  J155630$+$220729  & 0.682 p [5]    &  --             & Ext& [5]       \\
  J155644$+$220658  & 0.929 p [5]    &  --             & Ext& [5]       \\
  J160203$+$220931  & 0.555 p [5]    &  --             & Ext& [5]       \\
  J160317$+$215841  & 0.662 p [5]    &  Galaxy\,[2]     & Ext& [5]       \\
  J161105$+$220709  & --             &  --             & Ext &[5]        \\
  J161334$+$220425  & --             &  --             & EF &        \\
  J161423$+$220020  & 0.829 p [5]    &  --             & Ext& [5]       \\
  J161759$+$220136  & 0.418 p [5]    &  --             & Ext& [5]       \\
  J161847$+$215921  & 0.334 s [5]    & BL\,Lac\,[1]       & Pt& [5]        \\
  J162110$+$215739  & 0.689 p [5]    & --              & Ext& [5]       \\
  J162917$+$220628  & --             & --              & Ext& [5]       \\
  J164255$+$221226  & 0.834 p [5]    &  --             & Ext& [5]       \\
  J164439$+$220214  & 0.492 s [4]    &     QSO\,[4]     & Pt & [5]       \\
  J164631$+$215857  & 0.358 s [3]    & Galaxy\,[2]      & Ext& [5]            \\
		    & --             & QSO\,[3]          &    &     \\
  J164819$+$220114  & --             &                 & Pt& [5]        \\
  J170251$+$220532  & 0.535 p [5]    &  --             & Ext& [5]        \\
  J170744$+$220049  & 1.593 s [5]    &   QSO\,[2,5]     & Pt& [5]        \\
  J171332$+$215557  & 2.646 s [4]    &    QSO\,[4]      & Pt& [5]         \\
  J171611$+$215214  & 2.380 s [4]    &   FSRQ\,[1]     & Ext& [5]       \\
&& BL\,Lac [2]  &&\\
  J172003$+$215847  & 0.228 p [5]    &  --             & Ext& [5]        \\
  J172655$+$220102  & --             &  --             & Pt& [7]      \\
  J174005$+$221100  & 1.403 s [4,3]  &   QSO\,[2]        & Pt& [7]       \\
  J174306$+$215932  & --             &              & Ext& [3]      \\
  J174525$+$215703  & --             &   --            & Pt& [7]       \\
  J174536$+$220340  & --             &   --            & Pt& [7]      \\
  J175011$+$215734  & --             &   --            & Pt& [7]       \\
  J175915$+$215933  & --             &   --            &  VisS&      \\
  J180127$+$215732  & --             &   --            & Pt& [7]       \\
  J180738$+$220456  & 0.798 s [2]    &  QSO\,[2]        & Pt& [5]        \\
		    &                & FSRQ\,[1]        & &\\
  J181307$+$215430  &  --            &   QSO\,[5]       & Pt& [5]         \\
  J181725$+$215845  & --             &   --            & Pt& [5]        \\
  J182812$+$215519  & --             &   --            & Ext& [5]        \\
  J183118$+$220012  & 0.977 s [4]    &   QSO\,[2,4]     & Ext& [5]       \\
  J184035$+$215744  & --             &   --            & Pt& [7]       \\
  J184839$+$220118  & --             &   --            & Pt& [7]      \\
  J185423$+$215858  & --             &   --            & Pt& [7]       \\
  J191619$+$215719  & --             &   --            & Pt& [7]      \\
  J191640$+$220459  & --             &   --            & Pt& [7]       \\
  J193453$+$215701  & --             &   --            & Pt& [7]     \\
  J195455$+$215957  & --             &   --            & Pt& [7]       \\
  J195607$+$215941  & --             &   --            & Pt& [7]     \\
  J201311$+$220052  & --             &    QSO\,[4]      & Pt& [7]       \\
  J201553$+$215655  &  --            &   --            & Pt& [7]       \\
  J202545$+$215836  &   --           &    --           & Pt& [7] \\
  J203307$+$215905  & 2.413 s [4]    &   QSO\,[4]       & Pt& [7]     \\
  J203334$+$215934  & --             &   --            & Pt& [7]\\
  J203732$+$215303  &    --          &  --               & Pt & [7]       \\
  J203934$+$215209  & 2.782 s [4]    &   QSO\,[2,4]      & Pt(7)  \\
  J204007$+$215319  & --             &   --            & Pt& [7]     \\
  J205826$+$215819  & --             &   --            & Pt& [7]       \\
  J210908$+$215502  & 2.359 s [4]    &   QSO\,[2,4]      & Pt& [5]        \\
  J211032$+$215830  & --             &   --            & Pt& [7]       \\
  J212301$+$215047  & --             &   --            & Pt& [5]       \\
  J213057$+$214926  & --             &   --            & Pt& [7]       \\
  J213649$+$215701  & --             &   --            & VisS&    \\
  J213735$+$215738  & 0.193 p [5]    &   Galaxy\,[4]     & Pt &[5]         \\
  J214732$+$215434  & --             &   QSO\,[4]       & Ext& [5]      \\
  J221213$+$215521  &  0.584 p [5]   &   --            & Pt &[5]         \\
  J221828$+$215633  & --             &   --            & Ext& [5]      \\
  J222059$+$215222  & 0.375 p [5]    &   QSO\,[2]        & Ext &[5]        \\
  J222928$+$215435  & --             &   --            & Ext& [5]      \\
  J223659$+$215318  & 0.450 p [5]    &  --             & Ext& [5]        \\
  J224128$+$220019  & 0.537 p [5]    &  --             & Ext& [5]      \\
  J225825$+$215251  & 0.672 p [5]    &  --             & Pt & [5]        \\
  J231559$+$215435  & 1.170 s [5]    &   QSO\,[2,5]      & Ext& [5]      \\
  J232439$+$215548  & 0.438 p [5]    &  --             & Ext &[5]        \\
  J233611$+$215005  & --             &  --             & Pt &[5]       \\
  J233724$+$215847  & 2.218 s [5]    &   QSO\,[2,4,5]    & Pt& [5]         \\
  J233930$+$215630  & --             &  --             & Ext&  [5]     \\
  J234025$+$215509  & 0.416 s [5]    &   GinCl\,[2]      & Pt& [5]         \\
  J234516$+$215141  & 0.583 s [5]    &   QSO\,[2,4,5]    &  Pt& [5]      \\
  J234706$+$215251  & --             &  --             & EF& [5]         \\
  J234749$+$220016  & --             &  --             & Ext& [5]      \\
  J235240$+$215735  & 0.855 p [5]    &  --             & Ext& [5]        \\
  J235913$+$215732  & 0.704 s [5]    & Galaxy\,[5]       & Ext& [5]      \\
\hline

\end{longtable*}
\renewcommand{\baselinestretch}{1}

\subsection{ Radio luminosity of sources}

Radio luminosity at a frequency of 4.7 GHz was calculated for 112 objects with known redshifts. The calculations used
 $\Lambda$CDM cosmology with $H_0$ = 67.4 км\,с$^{-1}$\,Mpc$^{-1}$, $\Omega_m$~=~0.315 and $\Omega_\Lambda$~=~0.685~\citep{2020A&A...641A...6P}  to estimate radio luminosity using the formula
 \citep{1988gera.book..641C}:
\begin{equation}
L_{\nu} = 4 \pi D_{l}^2 \nu S_{\nu} (1+z)^{-\alpha -1},
\end{equation}
where $\nu$ is the frequency of observations, $S$ is the measured flux density at
the observation frequency, $z$ is the redshift, $\alpha$ is the spectral the index, $D_l$, is the luminosity distance (Hogg, 1999). Its values were obtained using the cosmology module of the {\tt astropy} package for the programming language
Python. The calculated values of radio luminosity at a frequency of 4.7 GHz and their errors are shown in table \ref{BursovA3}.
The accuracy of the radio luminosity estimate is determined by the errors of all values included in the equation. The large luminosity errors are related with significant photometric redshift errors.

\LTcapwidth=0.95\linewidth
\renewcommand{\baselinestretch}{0.73}
\setlength{\tabcolsep}{3.5pt}
\begin{longtable}
{c|c@{$\times$}rcr}
\caption{ The radio luminosity of the survey sources at a frequency of 4.7~ GHz. The calculated values of the luminosity with errors and relative errors of the radio luminosity $\delta L$ are indicated as a percentage
} \medskip
\label{BursovA3} \\
\hline
NVSS name    & \multicolumn{4}{l}{Luminosity, erg\,s$^{-1}$~($\delta L$)} \\
\endfirsthead
\caption{(Continued)\medskip  } \medskip\\
\hline
NVSS name    & \multicolumn{4}{l}{Luminosity, erg\,s$^{-1}$~~($\delta L$)}\\  \hline
\endhead
\hline
\endfoot
\endlastfoot
\hline
 J000727$+$220413   &   (1.0$\pm$0.1)&$ 10^{43}$& (10\%)          \\
 J001145$+$215912   &   (3.1$\pm$0.2)&$ 10^{43}$& (6\%)          \\
 J002130$+$215319   &   (1.9$\pm$0.3)&$ 10^{43}$& (16\%)          \\
 J002337$+$215624   &   (6.4$\pm$0.2)&$ 10^{41}$& (3\%)          \\
 J003147$+$215347   &   (4.6$\pm$1.0)&$ 10^{43}$& (22\%)          \\
 J004157$+$215423   &   (1.9$\pm$0.5)&$ 10^{42}$& (26\%)          \\
 J012428$+$215454   &   (1.7$\pm$0.1)&$ 10^{42}$& (6\%)          \\
 J012729$+$215136   &   (1.1$\pm$0.7)&$ 10^{42}$& (64\%)          \\
 J013352$+$220125   &   (2.5$\pm$0.3)&$ 10^{43}$& (12\%)          \\
 J013553$+$215816   &   (2.2$\pm$0.2)&$ 10^{44}$& (9\%)          \\
 J014235$+$215731   &   (1.1$\pm$0.08)&$10^{44}$& (7\%)         \\
 J015218$+$220707   &   (2.2$\pm$0.1)&$ 10^{44}$& (5\%)          \\
 J015641$+$215651   &   (3.1$\pm$1.5)&$ 10^{42}$& (48\%)          \\
 J020434$+$215328    &   (1.1$\pm$0.3)&$ 10^{42}$& (27\%)          \\
 J022110$+$215551    &   (1.7$\pm$0.8)&$ 10^{42}$& (47\%)          \\
 J022754$+$215451    &   (7.4$\pm$3.0)&$ 10^{40}$& (41\%)          \\
 J023001$+$215304    &   (3.0$\pm$1.4)&$ 10^{42}$& (47\%)          \\
 J023004$+$215909   &   (2.4$\pm$0.3)&$ 10^{42}$& (13\%)          \\
 J023349$+$215317   &   (1.0$\pm$0.4)&$ 10^{42}$& (40\%)          \\
 J041913$+$220304   &   (8.2$\pm$2.1)&$ 10^{42}$& (26\%)          \\
 J042211$+$220241   &   (5.4$\pm$1.2)&$ 10^{42}$& (22\%)          \\
 J043856$+$215157   &   (1.0$\pm$2.1)&$ 10^{42}$& (210\%)          \\
 J045004$+$215814   &   (3.8$\pm$2.4)&$ 10^{41}$& (63\%)          \\
 J072319$+$220100   &   (2.7$\pm$0.1)&$ 10^{40}$& (4\%)          \\
 J072351$+$220241   &   (6.8$\pm$0.3)&$ 10^{40}$& (4\%)          \\
 J072614$+$215319   &   (1.1$\pm$0.1)&$ 10^{44}$& (9\%)          \\
 J072820$+$215306   &   (1.9$\pm$0.2)&$ 10^{45}$& (11\%)          \\
 J081212$+$220024   &   (9.2$\pm$0.9)&$ 10^{42}$& (10\%)          \\
 J085037$+$220615   &   (2.9$\pm$0.3)&$ 10^{43}$& (10\%)          \\
 J090614$+$220010   &   (7.3$\pm$0.5)&$ 10^{42}$& (7\%)          \\
 J091224$+$220506   &   (2.6$\pm$0.3)&$ 10^{43}$& (12\%)          \\
 J091914$+$220519   &   (8.7$\pm$0.6)&$ 10^{43}$& (7\%)          \\
 J092045$+$220433   &   (4.4$\pm$4.4)&$ 10^{39}$& (100\%)          \\
 J092601$+$220136   &   (5.7$\pm$0.9)&$ 10^{42}$& (16\%)          \\
 J094736$+$220136   &   (1.6$\pm$1.0)&$ 10^{42}$& (63\%)          \\
 J094836$+$220053   &   (9.9$\pm$0.7)&$ 10^{42}$& (7\%)          \\
 J101401$+$215825   &   (8.6$\pm$6.4)&$ 10^{41}$& (74\%)          \\
 J102016$+$220940   &   (7.3$\pm$0.5)&$ 10^{41}$& (7\%)          \\
 J102154$+$215930   &   (5.6$\pm$2.1)&$ 10^{43}$& (38\%)          \\
 J102408$+$220347   &   (1.6$\pm$1.8)&$ 10^{42}$& (120\%)          \\
 J103633$+$220312   &   (1.2$\pm$0.1)&$ 10^{43}$& (8\%)          \\
 J103943$+$215743   &   (1.1$\pm$0.06)&$ 10^{43}$& (5\%)         \\
 J104254$+$220127   &   (3.7$\pm$0.6)&$ 10^{42}$& (16\%)          \\
 J104702$+$221033   &   (9.3$\pm$18.8)&$ 10^{39}$& (202\%)         \\
 J105430$+$221055   &   (1.1$\pm$0.2)&$ 10^{44}$& (18\%)          \\
 J110025$+$220156   &   (2.1$\pm$0.4)&$ 10^{42}$& (19\%)          \\
 J112829$+$220729   &   (7.2$\pm$2.1)&$ 10^{42}$& (29\%)          \\
 J113033$+$215728   &   (6.0$\pm$0.4)&$ 10^{42}$& (7\%)          \\
 J114325$+$220656   &   (1.4$\pm$0.04)&$ 10^{44}$& (3\%)         \\
 J114417$+$220752   &   (2.8$\pm$0.2)&$ 10^{42}$& (7\%)          \\
 J114821$+$220825   &   (1.1$\pm$0.4)&$ 10^{43}$& (36\%)          \\
 J115311$+$220654   &   (9.4$\pm$5.0)&$ 10^{41}$& (53\%)          \\
 J121156$+$220455   &   (3.0$\pm$2.1)&$ 10^{40}$& (70\%)          \\
 J125433$+$221103   &   (2.0$\pm$0.2)&$ 10^{42}$& (10\%)          \\
 J130253$+$220758   &   (4.1$\pm$2.9)&$ 10^{41}$& (71\%)          \\
 J130651$+$221119   &   (2.4$\pm$2.4)&$ 10^{42}$& (100\%)          \\
 J131128$+$220306   &   (7.3$\pm$4.3)&$ 10^{41}$& (59\%)       \\
 J132700$+$221050   &   (1.9$\pm$0.1)&$ 10^{44}$& (5\%)       \\
 J132749$+$220503   &   (3.1$\pm$0.2)&$ 10^{43}$& (6\%)       \\
 J133212$+$220549   &   (2.2$\pm$0.4)&$ 10^{42}$& (18\%)       \\
 J133928$+$220822   &   (3.6$\pm$0.4)&$ 10^{43}$& (11\%)       \\
 J135116$+$221110   &   (2.7$\pm$0.3)&$ 10^{43}$& (11\%)       \\
 J135313$+$220540   &   (1.2$\pm$0.9)&$ 10^{42}$& (75\%)       \\
 J140808$+$220155   &   (3.3$\pm$1.2)&$ 10^{42}$& (36\%)       \\
 J141242$+$215939   &   (1.7$\pm$0.8)&$ 10^{42}$& (47\%)       \\
 J141351$+$220647   &   (4.7$\pm$4.2)&$ 10^{41}$& (89\%)       \\
 J141726$+$220539   &   (2.8$\pm$0.6)&$ 10^{42}$& (21\%)       \\
 J143106$+$220505   &   (2.4$\pm$0.6)&$ 10^{42}$& (25\%)       \\
 J144057$+$220142   &   (2.4$\pm$1.5)&$ 10^{40}$& (63\%)       \\
 J150123$+$221122   &   (5.5$\pm$2.1)&$ 10^{42}$& (38\%)       \\
 J151105$+$220806   &   (1.3$\pm$0.8)&$ 10^{43}$& (62\%)       \\
 J151319$+$220255   &   (6.0$\pm$2.4)&$ 10^{42}$& (40\%)       \\
 J151830$+$220313   &   (3.6$\pm$1.3)&$ 10^{42}$& (36\%)       \\
 J153652$+$220207   &   (1.9$\pm$0.3)&$ 10^{41}$& (16\%)       \\
 J154535$+$220400   &   (1.5$\pm$0.7)&$ 10^{41}$& (47\%)       \\
 J154631$+$215741   &   (1.8$\pm$0.1)&$ 10^{43}$& (6\%)       \\
 J155354$+$215927   &   (2.4$\pm$0.1)&$ 10^{43}$& (4\%)       \\
 J155630$+$220729   &   (4.6$\pm$1.9)&$ 10^{42}$& (41\%)       \\
 J155644$+$220658   &   (8.4$\pm$1.9)&$ 10^{42}$& (23\%)       \\
 J160203$+$220931   &   (4.9$\pm$3.0)&$ 10^{42}$& (61\%)       \\
 J160317$+$215841   &   (4.0$\pm$1.6)&$ 10^{42}$& (40\%)       \\
 J161423$+$220020   &   (7.7$\pm$1.7)&$ 10^{42}$& (22\%)       \\
 J161759$+$220136   &   (4.4$\pm$3.6)&$ 10^{41}$& (82\%)       \\
 J161847$+$215921   &   (5.7$\pm$0.5)&$ 10^{41}$& (9\%)       \\
 J162110$+$215739   &   (2.8$\pm$1.9)&$ 10^{42}$& (68\%)       \\
 J164255$+$221226   &   (1.6$\pm$0.4)&$ 10^{43}$& (25\%)       \\
 J164439$+$220214   &   (3.6$\pm$0.2)&$ 10^{42}$& (6\%)       \\
 J164631$+$215857   &   (7.3$\pm$0.5)&$ 10^{41}$& (7\%)       \\
 J170251$+$220532   &   (2.1$\pm$0.5)&$ 10^{42}$& (24\%)       \\
 J170744$+$220049   &   (1.5$\pm$0.2)&$ 10^{43}$& (13\%)       \\
 J171332$+$215557   &   (6.7$\pm$0.5)&$ 10^{43}$& (7\%)       \\
 J171611$+$215214   &   (4.5$\pm$0.3)&$ 10^{44}$& (7\%)       \\
 J172003$+$215847   &   (4.6$\pm$7.2)&$ 10^{41}$& (157\%)       \\
 J174005$+$221100   &   (7.2$\pm$1.0)&$ 10^{43}$& (14\%)       \\
 J180738$+$220456   &   (9.6$\pm$0.6)&$ 10^{42}$& (6\%)       \\
 J183118$+$220012   &   (3.5$\pm$0.2)&$ 10^{43}$& (6\%)       \\
 J203307$+$215905   &   (1.5$\pm$0.1)&$ 10^{44}$& (7\%)       \\
 J203934$+$215209   &   (3.0$\pm$0.1)&$ 10^{44}$& (3\%)       \\
 J210908$+$215502   &   (1.3$\pm$0.07)&$10^{44}$& (5\%)      \\
 J213735$+$215738   &   (2.1$\pm$1.1)&$ 10^{41}$& (52\%)       \\
 J221213$+$215521   &   (2.5$\pm$0.5)&$ 10^{42}$& (20\%)       \\
 J222059$+$215222   &   (5.7$\pm$2.5)&$ 10^{41}$& (44\%)       \\
 J223659$+$215318   &   (9.3$\pm$1.5)&$ 10^{41}$& (16\%)       \\
 J224128$+$220019   &   (2.0$\pm$0.9)&$ 10^{42}$& (45\%)       \\
 J225825$+$215251   &   (1.5$\pm$0.9)&$ 10^{42}$& (60\%)       \\
 J231559$+$215435   &   (1.5$\pm$0.1)&$ 10^{43}$& (7\%)       \\
 J232439$+$215548   &   (1.5$\pm$1.3)&$ 10^{42}$& (87\%)       \\
 J233724$+$215847   &   (5.6$\pm$0.6)&$ 10^{43}$& (11\%)       \\
 J234025$+$215509   &   (6.8$\pm$0.6)&$ 10^{41}$& (9\%)       \\
 J234516$+$215141   &   (4.3$\pm$0.3)&$ 10^{42}$&  (7\%)       \\
 J235240$+$215735   &   (6.3$\pm$1.6)&$ 10^{42}$&  (25\%)       \\
 J235913$+$215732   &   (3.3$\pm$0.3)&$ 10^{42}$&  (9\%)       \\
\hline
\end{longtable}
\renewcommand{\baselinestretch}{1.0}

\begin{figure}[t]
\includegraphics[width=0.49\textwidth]{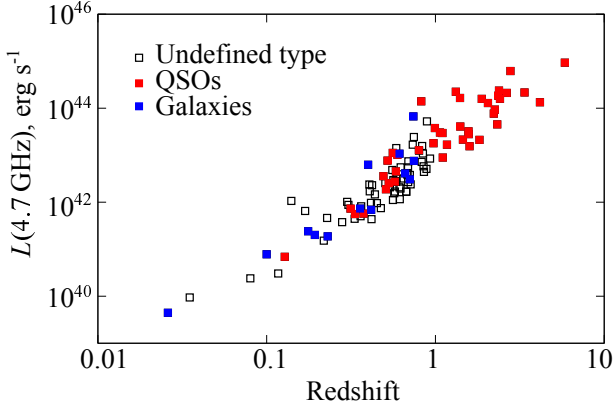}
\caption{Luminosity of sources at a frequency of \mbox{4.7~GHz} depending on the redshift for quasars, galaxies and sources of unknown type.}
\label{ris:lum}
\end{figure}

Figure\ref{ris:lum}  shows the distribution of calculated values of the radio luminosity of sources at a frequency of 4.7 GHz depending on the redshift for quasars, galaxies and sources of unknown type. Considering that
objects with low luminosity are mostly extended objects, it can be assumed that these objects are galaxies. The objects with the highest luminosity are distant quasars.

\section{CONCLUSION}

1. In the 2018-2019 survey, 205 bright sources were detected in the Western Sector of the RATAN-600 radio telescope at the declination of the pulsar in the Crab Nebula ($\rm Dec=22^\circ$) on a three-beam radiometric complex at a frequency of 4.7 GHz. The average flux densities were determined for all sources: for a year of observations, for each of the 12 months and for every three days for the 26 brightest of them.

2. According to the data obtained in this work, as well as from the CATS database,  
spectra of all 205 objects in the sample are plotted. 124 sources (61\%) have a power spectrum 
($-1.1 < \alpha < -0.5$, \mbox{$S_\nu \sim \nu^\alpha$}), 22 sources with peak distribution in the radio spectrum are 
$CSS -- GPS$ sources with a maximum radiation of 0.1-5 GHz and are assumed to belong to young compact objects; 25 sources are sources from the USS ($\alpha  < -1.1$) and probably related to distant radio galaxies in clusters with active star-formation, HzRGs, or compact objects of the CSS type; 26 sources have a flat spectrum ($-0.5 < \alpha  < 0.3$) and are quasars and blazars; three sources have an inverted spectrum ( $\alpha  > $ 0.3), possibly thermal; eight sources have a rise in the spectrum at low frequencies. For a quarter of the sources in the high-frequency region of the spectrum ($\nu > 4$ GHz), the data were obtained for the first time, and for the rest they were supplemented or refined. The mediate value of the spectral index distribution is  $\alpha \approx -0.9$.

3. Depending on the “flux density – spectral index” the twisting of spectral indices is traced with a decrease in the density of the source flux, and depending on "redshift – spectral index" is the flattening of spectral indices with increasing z, which is inconsistent with such distributions at low frequencies. We believe that this is due to the significant contribution of distant quasars with powerful radio emission.

4. Estimates of the variability of the source are obtained, both according to the index of variability ($I_{\rm var}$) at intervals of one year and at an interval of three days for the 26 brightest sources. The  $I_{\rm var}$ assessment showed that most of the sample sources (97\%) had no significant change ($I_{\rm var} < 0.15$) in the flux density. However, blazar B2 1324+224 showed a twofold increase in the flux density within one year (Ivar = 0.3). At the brightest viewing source  (\mbox{$S_{1.4} > 1$}) NVSS J060351+215937 (4C +22.12), a candidate for blazars, no change in the radiation level was detected. Among the sources studied for daily variability, two (NVSS J043458+21555 and NVSS J161334+22042) are located in empty SDSS fields. For seven more sources from the literature, their class is unknown.

5. All  detected sources are identified with optical catalog sources SDSS (DR16), Gaia (DR3) Extragalactic, Pan-
STARRS, Guide Star Catalog, Infrared 2MASS and with the SIMBAD, NED, Roma-BZCAT databases sources. Among the sample objects quasars are 18\%, blazars — 6\%, galaxies — 6\%, in empty fields — 6\%.  28\% of sources belong to point objects, 34\% are extended ones. Extended objects are likely associated with the fainter or distant galaxies. Five objects have been identified only with infrared data. The redshift is known for 112 identified objects (57\%). Luminosity has been calculated for these sources, the largest of which is shown by distant quasars.

\section*{Acknowledgements}

The observational data were obtained at the unique scientific installation of the radio telescope RATAN-600 SAO RAS. \\
Observations on the telescopes of the SAO RAS are carried out with the support of the Ministry of Science and Higher Education of the Russian Federation. \\
The instrument base is being updated within the framework of the national project “Science and Universities". \\
The study was performed using the NASA/IPAC Extragalactic Database (https://ned.ipac.caltech.edu /); CATS database, available on the website of the Special Astrophysical Observatory of the Russian Academy of Sciences; SIMBAD database, operating in CDS, Strasbourg, France. The tools for accessing the VizieR catalog, CDS, Strasbourg , France, were used. \\
The authors are grateful to the reviewer for constructive comments that contributed to the improvement of the article, and to the editorial board of the journal for technical support in the process of its preparation.

\onecolumngrid

\section*{Radio Spectra}
\includegraphics[width=145 mm]{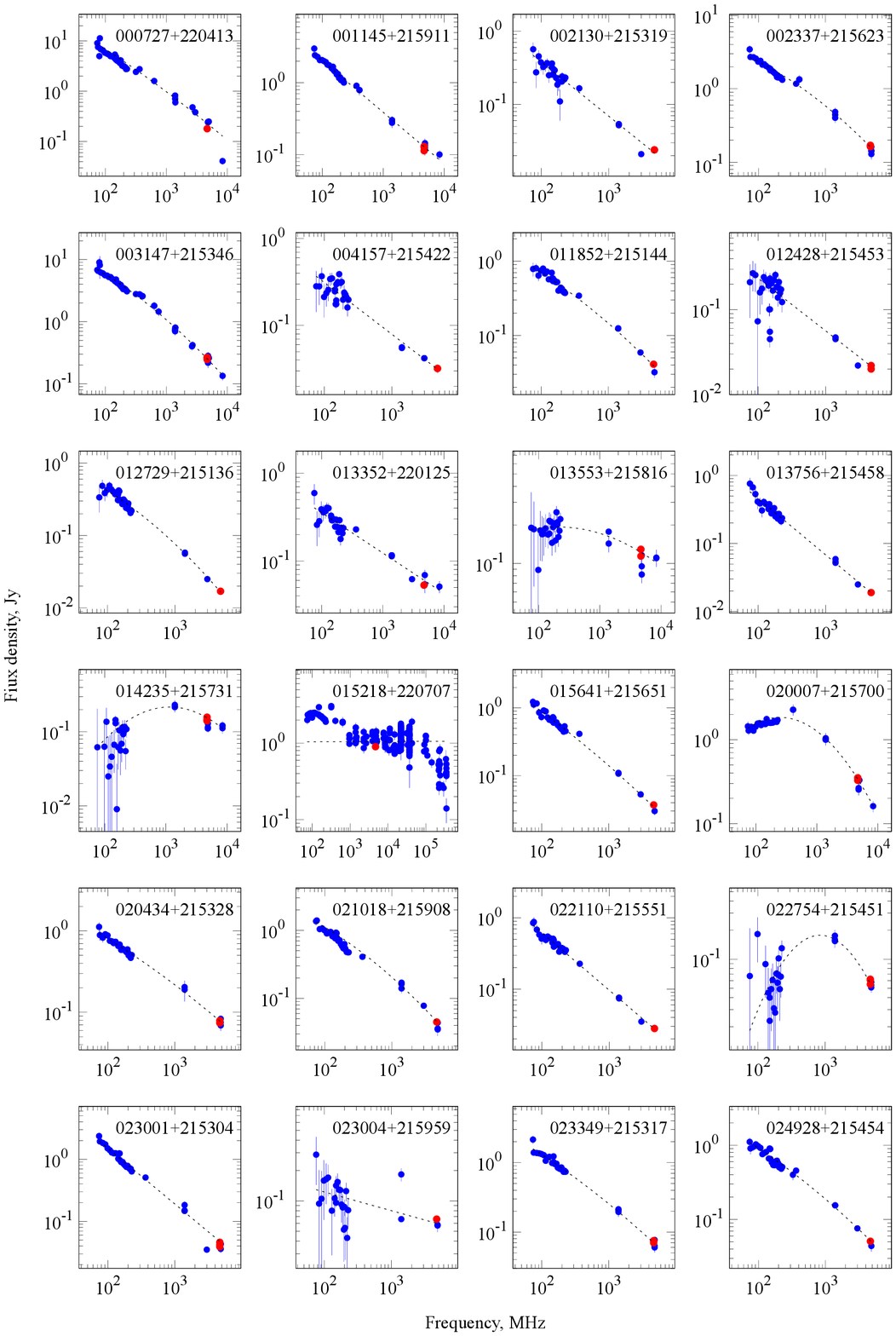}
\clearpage
\includegraphics[width=145 mm]{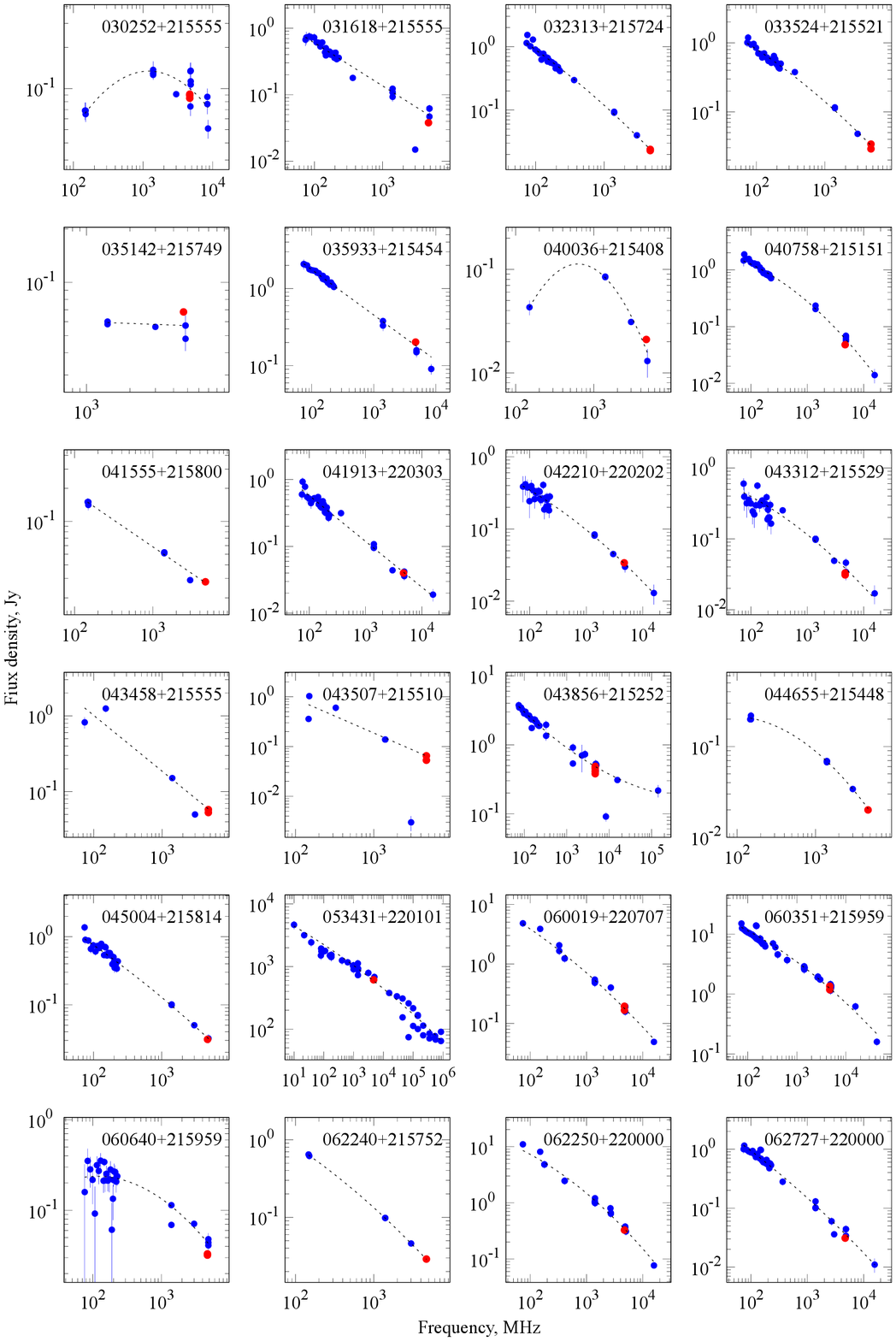}
\clearpage
\includegraphics[width=145 mm]{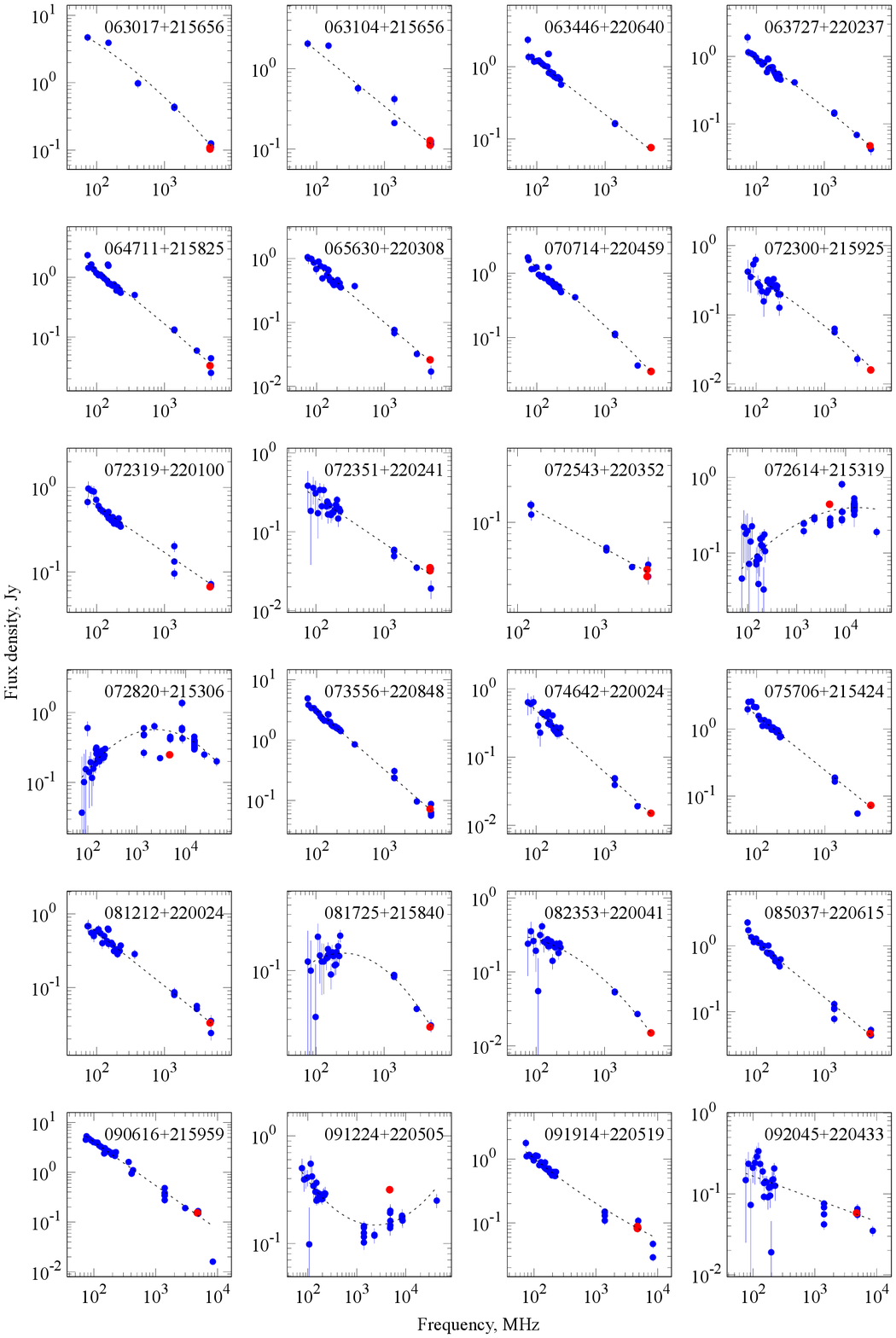}
\clearpage
\includegraphics[width=145 mm]{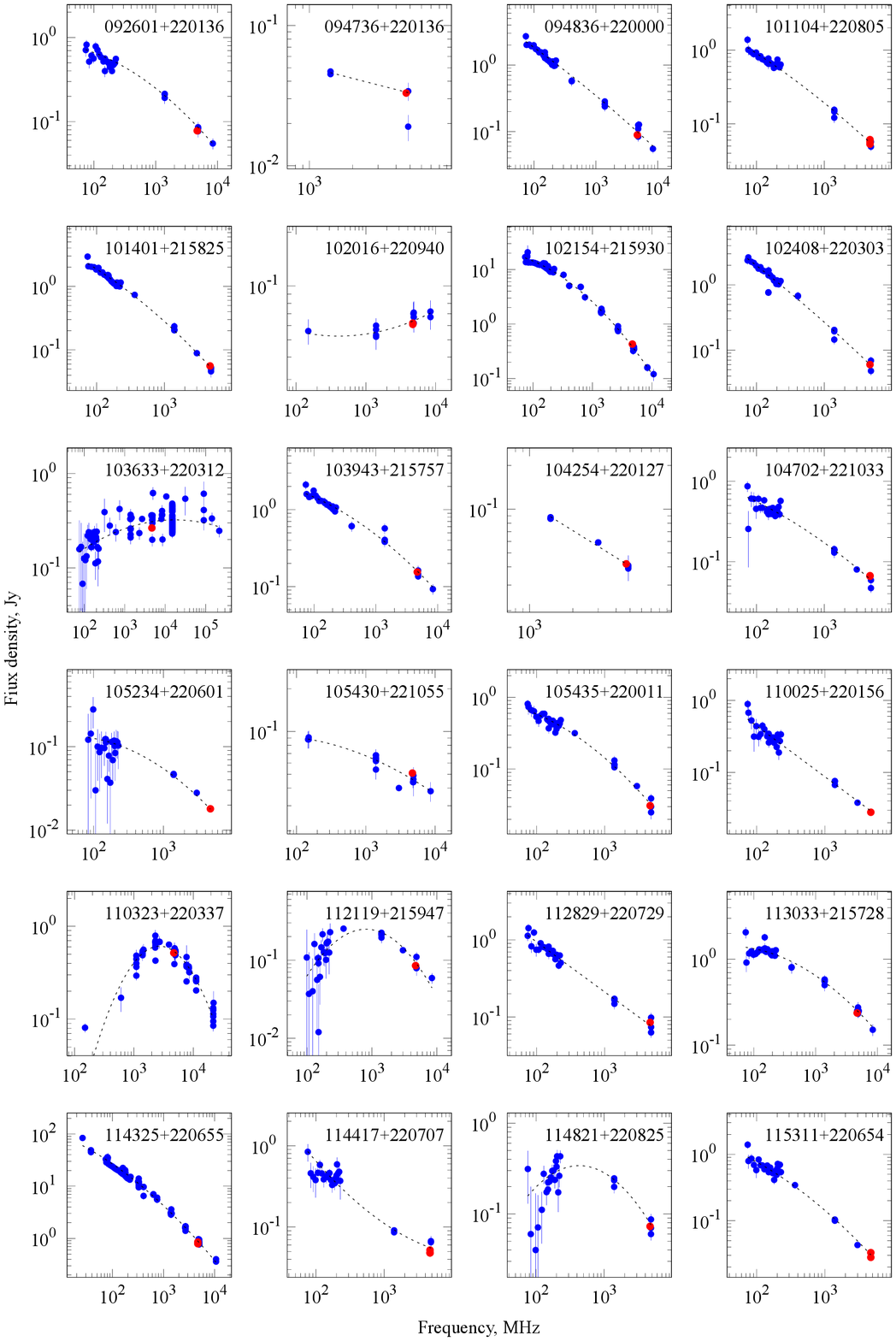}
\clearpage
\includegraphics[width=145 mm]{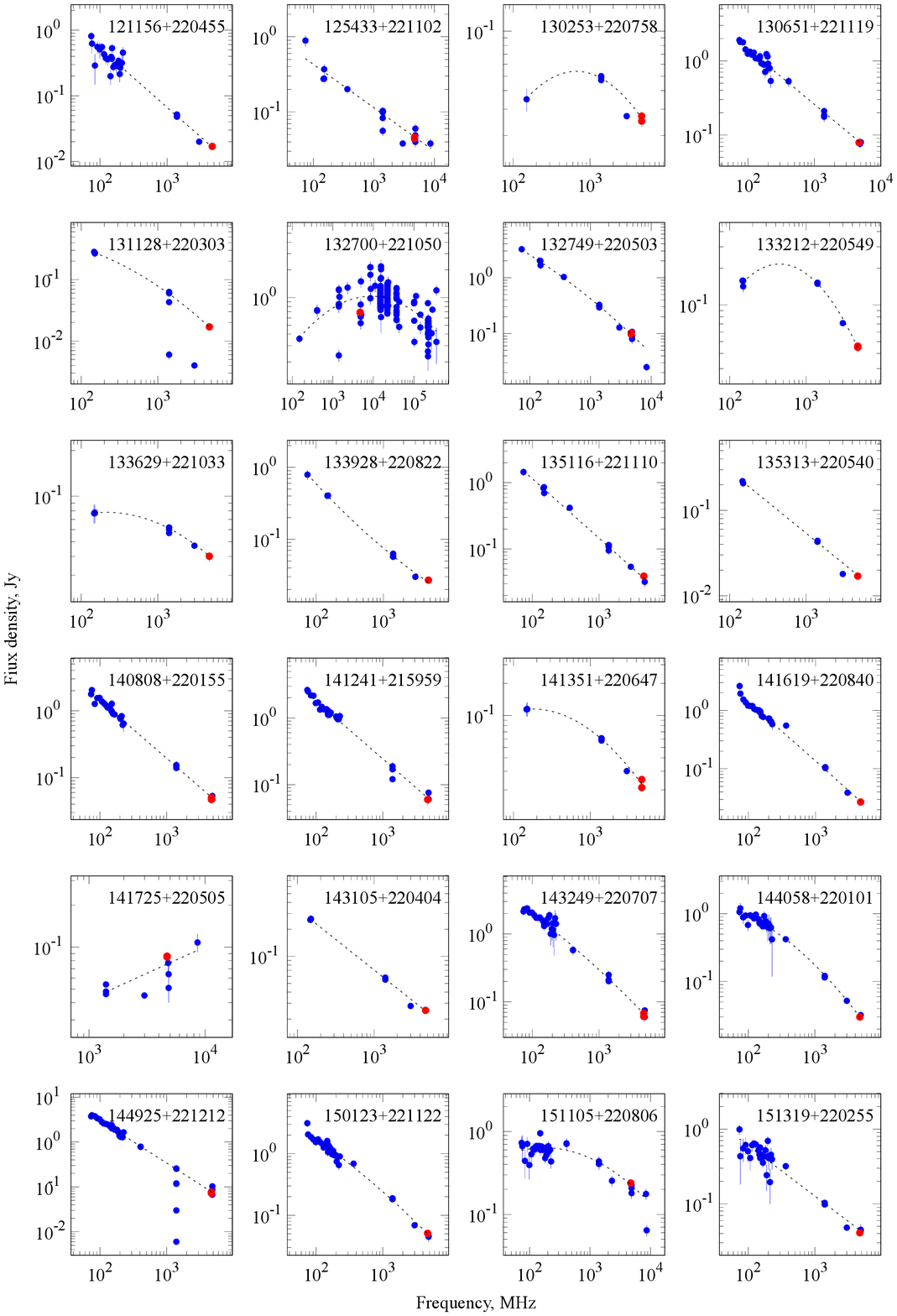}
\clearpage
\includegraphics[width=145 mm]{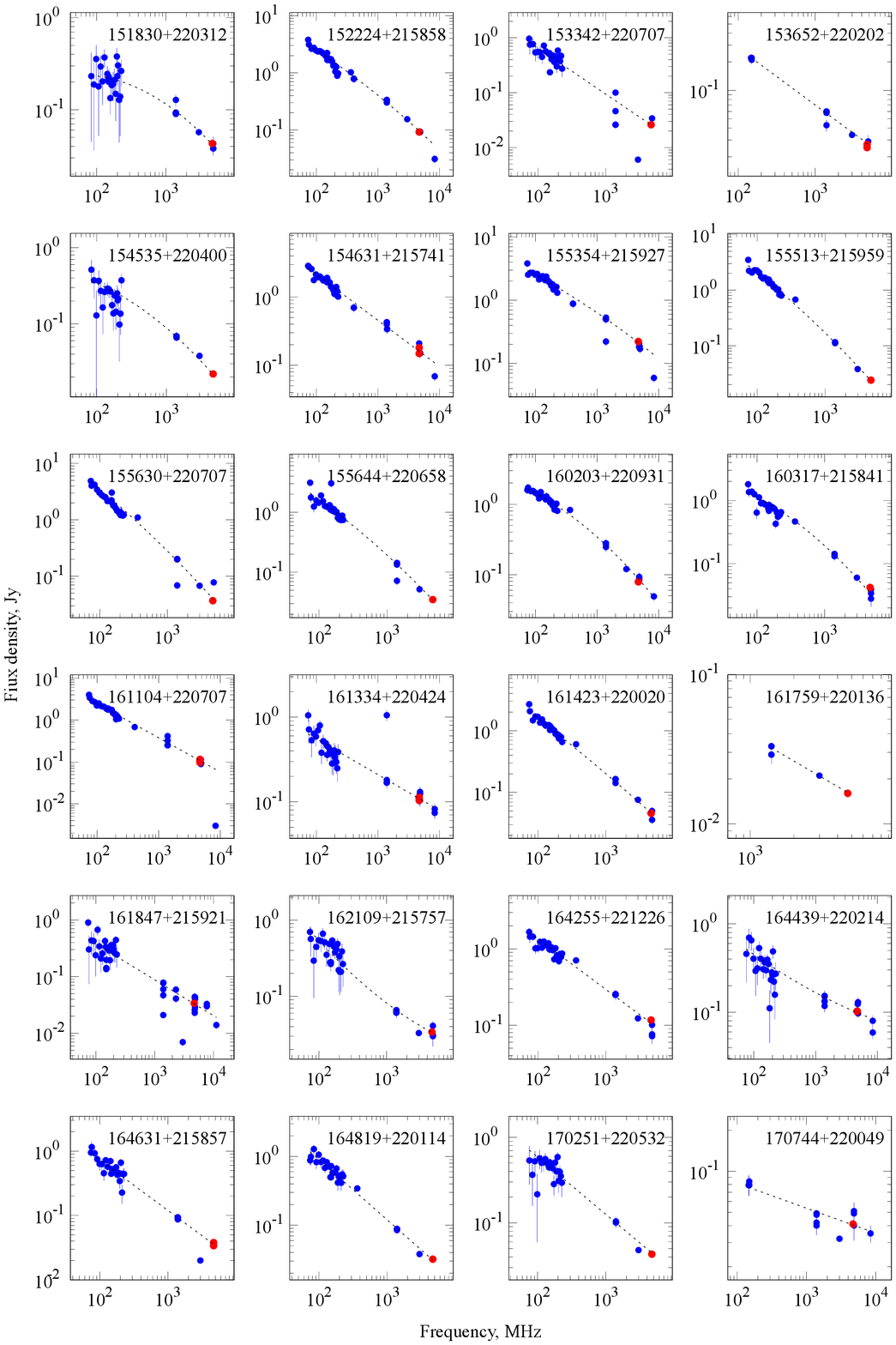}%
\clearpage
\includegraphics[width=145 mm]{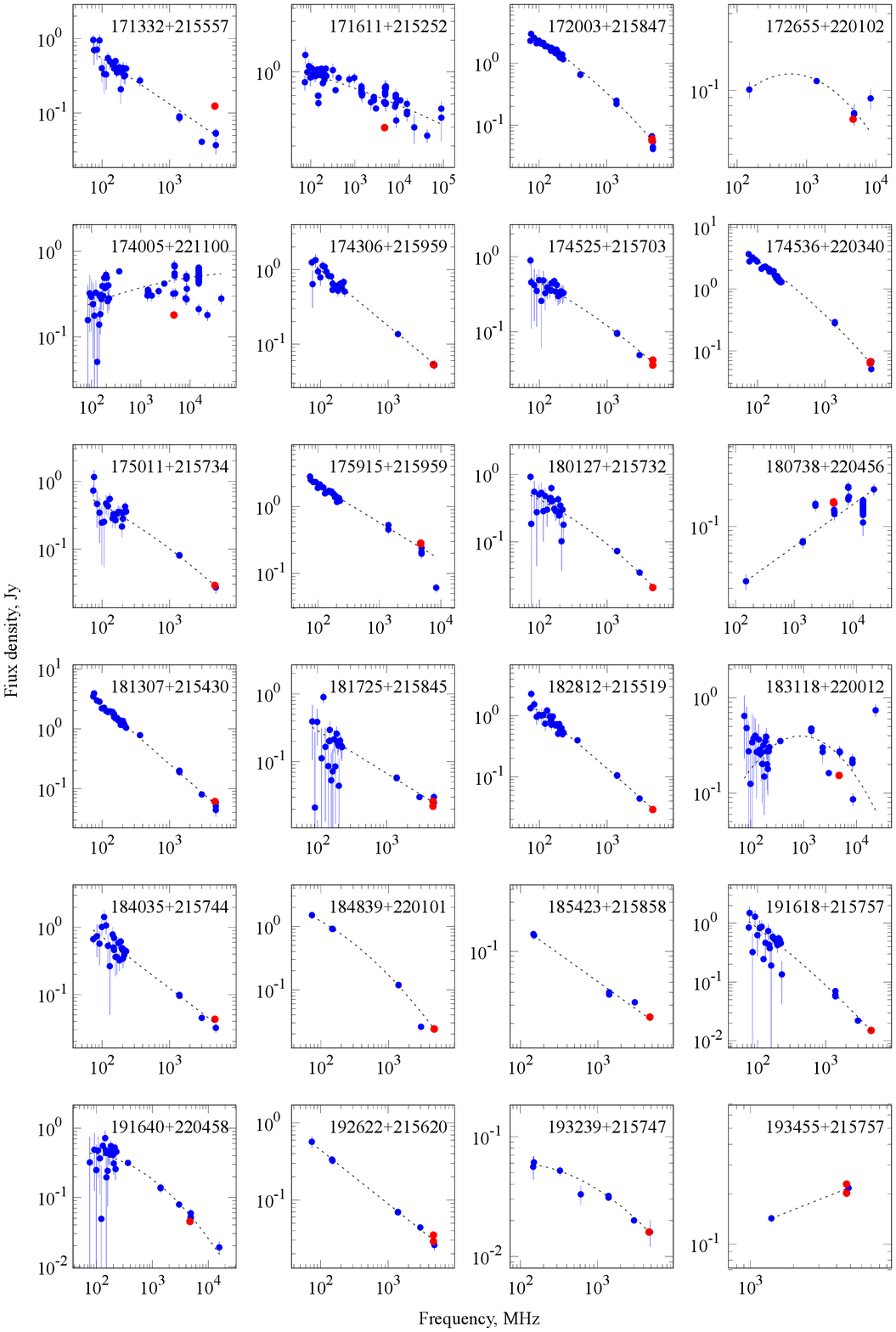}
\clearpage
\includegraphics[width=145 mm]{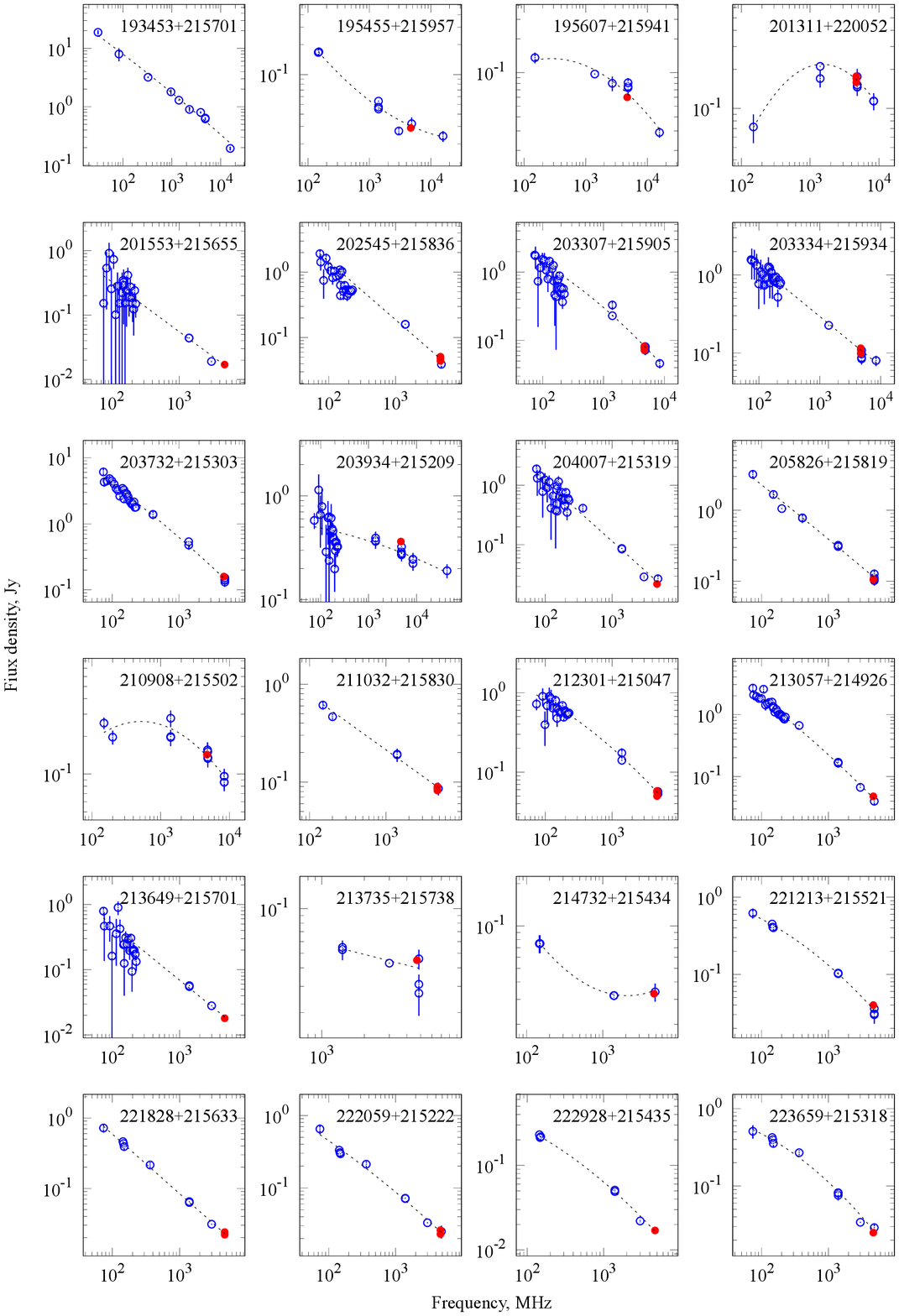}
\clearpage
\includegraphics[width=145mm]{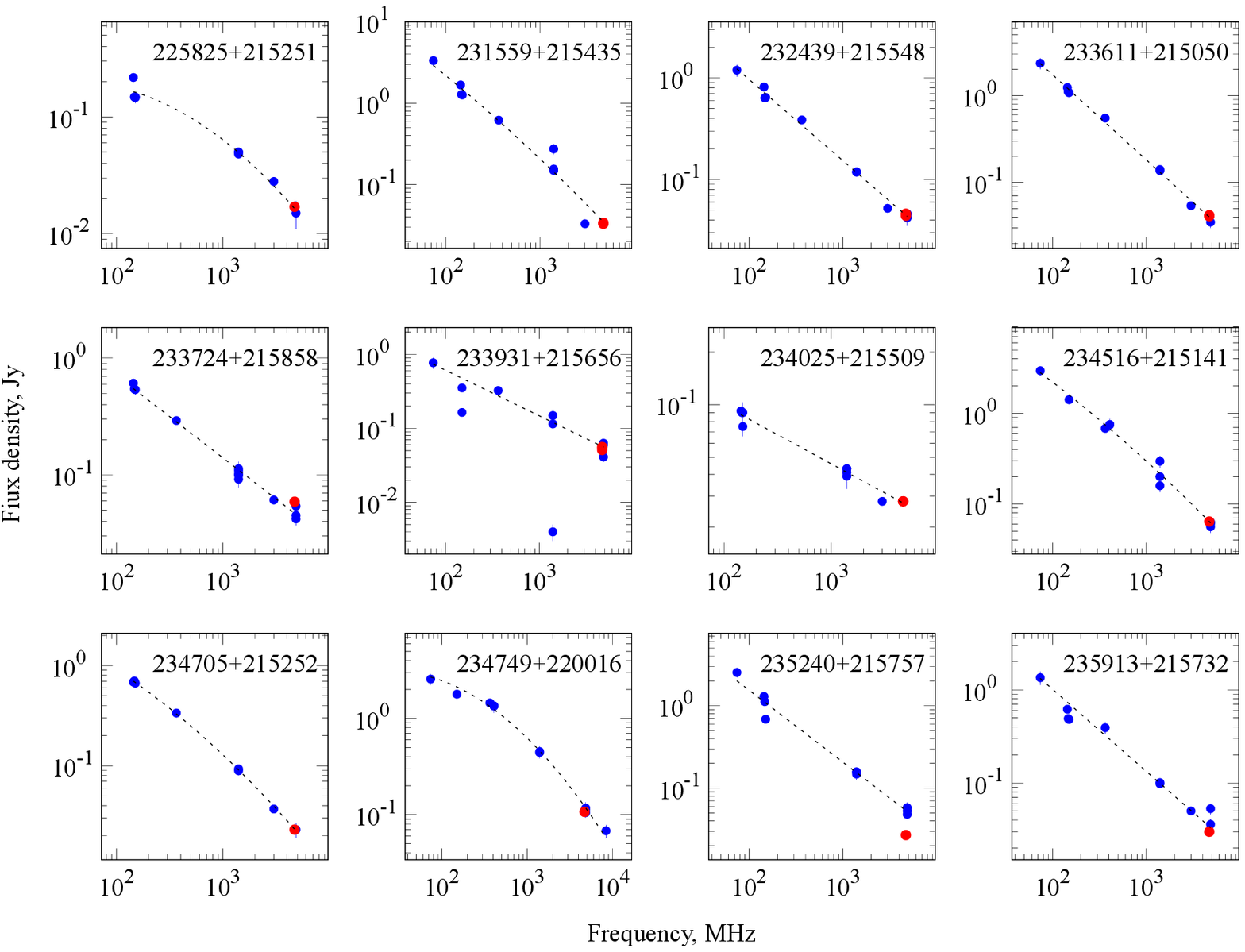}
\clearpage
\end{document}